\documentclass[a4paper,fleqn,usenatbib]{mnras}

\usepackage{newtxtext,newtxmath}
\usepackage{txfonts}

\usepackage[T1]{fontenc}
\usepackage{ae,aecompl}

\usepackage{graphicx}	
\usepackage{amssymb}	

\usepackage{savesym}
\usepackage{amsmath}	
\savesymbol{iint}
\usepackage{txfonts}
\restoresymbol{TXF}{iint}

\title[Thorium in solar twins]{Thorium in solar twins: implications for habitability in rocky planets}

\author[R. B. Botelho et al.]{
R. B. Botelho$^{1}$,
A. de C. Milone$^{1}$\thanks{E-mail: andre.milone@inpe.br (ACM)},
J. Melendez$^{2}$,
M. Bedell$^{3}$,
L. Spina$^{4}$, 
\newauthor M. Asplund$^{5}$, L. dos Santos$^{6}$, J. L. Bean$^{7}$, I. Ramirez$^{8}$, D. Yong$^{5}$, S. Dreizler$^{9}$, 
\newauthor A. Alves-Brito$^{10}$ and J. Yana Galarza$^{2}$
\\
$^{1}$Divis\~ao de Astrof\'\i sica, Instituto Nacional de Pesquisas Espaciais,
Av. dos Astronautas 1758 S\~ao Jos\'e dos Campos, 12227-010, Brazil\\
$^{2}$Departamento de Astronomia, IAG, Universidade de S\~ao Paulo, Rua do Mat\~ao 1226 S\~ao Paulo, 05509-900, Brazil\\
$^{3}$Center for Computational Astrophysics, Flatiron Institute, 162 5$^{\rm th}$ Ave, New York, NY 10010, USA\\
$^{4}$Monash Centre for Astrophysics, School of Physics and Astronomy, Monash University, VIC 3800, Australia\\
$^{5}$Research School of Astronomy and Astrophysics, The Australian National University,
Cotter Road, Canberra, ACT 2611, Australia\\
$^{6}$Observatoire de l'Universit\'e de Gen\'eve, 51 chemin des Maillettes, CH-1290 Versoix, Switzerland\\
$^{7}$Department of Astronomy \& Astrophysics, 5640 S. Ellis Ave, Chicago, IL 60637, USA\\
$^{8}$Tacoma Community College, 6501 South 19th Street, Tacoma, WA 98466-7400, USA\\
$^{9}$Institut f\"{u}r Astrophysik, Georg-August Universit\'at G\"{o}ttingen,
Wilhelmsplatz 1, D-37073, G\"{o}ttingen, Germany\\
$^{10}$Universidade Federal do Rio Grande do Sul, Instituto de F\'\i sica,
Av. Bento Gon\c calves 9500, Porto Alegre, RS, Brazil\\
}

\date{Accepted XXX. Received YYY; in original form ZZZ}

\pubyear{2018}

\begin{document}
\label{firstpage}
\pagerange{\pageref{firstpage}--\pageref{lastpage}}
\maketitle

\begin{abstract}

We have investigated the thorium (Th) abundance in a sample of 53 thin disc solar twins covering 
a wide range of ages. These data provide constrains on
the mantle energy budget of terrestrial planets
that can be formed over the evolution of the Galaxy's thin disc.
We have estimated Th abundances with an average precision of 0.025\,dex (in both [Th/H] and [Th/Fe])
through comprehensive spectral synthesis of a Th\,II line present at 4019.1290\,{\AA}, 
using very high resolution (R\,=\,115,000) high quality HARPS spectra obtained
at the ESO La Silla Observatory.
We have confirmed that there is a large energy budget from Th decay
for maintaining mantle convection inside potential rocky planets around solar twins,
from the Galactic thin disc formation until now,
because the pristine [Th/H]$_{\rm ZAMS}$ is super-solar on average
under a uniform dispersion of 0.056\,dex
(varying from +0.037 up to +0.138\,dex based on linear fits against isochrone stellar age).
Comparing to neodymium (Nd) and europium (Eu), two others neutron-capture elements,
the stellar pristine abundance of Th follows Eu along the Galactic thin disc evolution,
but it does not follow Nd,
probably because neodymium has a significant contribution from the $s$-process (about 60\,per\,cent).

\end{abstract}




%
\section{Introduction}
\label{introduction}

The structure of a terrestrial planet is driven by the formation process itself based on
the collision and accretion of planetesimals with differential mineralogical settling 
and also by the internal heat budget.
As the prototype geologically dynamic planet,
Earth has a partially-crystalized metallic dynamo core, and
a silicate liquid mantle, whose convection maintains plate tectonics of a rocky crust.
The major internal heat sources of Earth comes from
the secular cooling of core and mantle,
and the radioactive decay of Th, U and K isotopes in the mantle \citep{Huang2013}.
The mantle thickness and convection inside telluric planets
are directly linked to the abundances of these isotopes in the mantle
(basically $^{232}$Th, $^{235}$U and $^{238}$U) \citep{McDonough2003}.
The volcanic activity generated by the plate tectonics recycles gases
such as carbon dioxide between the atmosphere and the mantle,
contributing to habitability by keeping the planet surface at moderate temperatures
\citep{Walker1981}.
A geologically active planet could even play a role for the origin of life and its long-term maintenance
(e.g. \citet{Misra2015} and references therein).

\citet{Unterborn2015}
were the first to speculate about the potential energy budget of terrestrial planets
directly connected to the stellar photospheric abundance of thorium (Th).
They measured log\,$\epsilon$(Th)\footnote{log\,$\epsilon$(X)\,=\,log\,($n$(X)/$n$(H))\,+\,12, where $n$ is number density.}
varying from 59 up to 251\,per\,cent of the solar value in a sample of 13 solar twins and analogues,
indicating that possible terrestrial planets formed around these stars
have sufficient internal energy budget to sustain a convective mantle in a dynamic planet
and, consequently, increases the probability of having a habitable surface.
They also found that the Sun is depleted in Th in comparison with 13 solar twins/analogues.
They argued that a large variation in Th abundance could be explained by differences in stellar age within their sample,
besides other main possible factors such as
the continuum normalization of the analysed blend Fe-Ni-Mn-Th-Co-CN-Ce-CH at 4019\,{\AA}
(covering $\lambda\lambda$4018.9--4019.2\,{\AA})
and the difference in spectral resolution between two data sets (R\,=\,48,000 and R\,=\,115,000).
Specifically, they did not report any deep analysis of the Th abundance as a function of stellar age
(taken from \citet{Baumann2010}).
They concluded that the observed large Th abundance variation is likely due to 
primordial chemical inhomogeneity among the stars.

Thorium is a $r$-process element synthesized in core-collapse supernovae,
neutron star mergers and neutron star\,-\,black hole mergers in the Galaxy
($r$-process elements are those built up by rapid capture of neutrons into nuclei,
i.e. over energetic processes in which the neutrons flux is high enough to become negligible the effect of the free neutrons decay).
The half time of its most abundant radioactive isotope $^{232}$Th
(14.05\,Gyr) is as high as the age of Universe \citep{Cayrel2001}.
Previous works have found a large variation in the abundance of $r$-process elements (Th included)
covering both metal-poor old halo stars and the younger more metal-rich disc stars
\citep{delPeloso2005a, delPeloso2005b, Honda2004, JohnsonBolte2001}.
\citet{Butcher1987} was the pioneer in estimating the age of the Galaxy by measuring Th abundance in G-dwarfs,
taking the advantage of its sizable decay during the lifetime of the Galaxy. 
How the abundance of Th has evolved during the Galactic disc's life is still an open question
-- more specifically in the thin disc at the Sun's galactocentric distance.
Nearby solar twins spanning a range of ages can provide clues about this question
because their photospheric abundances reflect the composition of the interstellar medium at the time they formed.
As planets and stars are likely formed from the same natal cloud,
potential terrestrial planets should have a Th abundance related to the Th abundance of its host star.

We have homogeneously measured Th abundances in a sample of nearby solar twins (including two planet-host stars)
of the Galactic thin disc, covering a broad range of stellar ages,
in order to provide constrains about the mantle energy budget of terrestrial planets
during the evolution of the thin disc.
A Th\,II line present in a multi-species blend at 4019\,{\AA}
has been carefully analysed for this purpose, through comprehensive spectral synthesis differentially to the Sun.
Section~\ref{sample} presents the sample of solar twins, their main parameters, and the spectroscopic data.
In Section~\ref{methodology} we describe the spectral synthesis approach employed for measuring Th abundance in a multi-species blend.
Section~\ref{results} covers the analysis of resulting Th abundance ratios (current and primordial) against metallicity and stellar age.
Finally, we show the concluding remarks in Section~\ref{conclusions}.

%
\section{Sample of solar twins and observational data}
\label{sample}

The sample is composed of 67 solar twins (stars with effective temperature, surface gravity and metallicity around the solar values
within $\pm$\,100\,K in $T_{\rm eff}$ and within $\pm$\,0.1 dex in log\,$g$ and [Fe/H]),
which were recently analysed by \citet{Spina2018} and \citet{Bedell2018}.
Their spectroscopic data and main parameters adopted are described in this section.

\citet{Spina2018} derived their photospheric parameters
(and abundance of 12 neutron-capture, $n$-capture, elements)
by applying a line-by-line differential spectroscopic analysis relative to the Sun
through equivalent width (EW) measurements of Fe\,I and Fe\,II lines.
The estimated typical errors in $T_{\rm eff}$, log\,$g$, [Fe/H] and $\xi$
(micro-turbulence velocity) are, respectively, 4\,K, 0.012, 0.004\,dex and 0.011\,km\,s$^{-1}$.
We have also adopted the stellar masses $m$ and isochrone ages $t$ as derived by \citet{Spina2018},
which are represented by the more probable values
(i.e. that corresponds to the peak of a probability distribution)
by applying the isochrone method with the {\bf ${\rm q}^{2}$} code of \citet{Ramirez2014a, Ramirez2014b}.
Only for the two very young stars HIP\,3203 and HIP\,4909, whose more probable age values are null,
the ages are better represented by the average values,
as also adopted by our previous works \citep{Spina2018, Bedell2018}.
The typical age uncertainty is 0.4\,Gyr, which is derived from the probability distribution itself.
Besides the 12 $n$-capture elements (Sr, Y, Zr, Ba, La, Ce, Pr, Nd, Sm, Eu, Gd and Dy),
we have also considered the abundances of 17 additional elements analysed by \citet{Bedell2018}
(C, O, Na, Mg, Al, Si, S, Ca, Sc, Ti, V, Cr, Mn, Co, Ni, Cu and Zn).
The aforementioned works measured elemental abundances by a strict line-by-line EW analysis relative to the Sun.

HARPS (High Accuracy Radial velocity Planet Searcher) spectra
are used in this work to extract Th abundances.
HARPS is an ultra-stable echelle spectrograph installed on the 3.6m telescope
of the European Southern Observatory (ESO) at La Silla Observatory in Chile \citep{Mayor2003}.
The HARPS spectra covers $\lambda\lambda$3780--6910\,{\AA} under a resolving power R\,=\,115,000.
The spectra adopted in the current work are those analysed by \citet{Spina2018, Bedell2018}.
Each one-dimensional spectrum comes from more than 50 stacked spectra
that are previously Doppler-corrected and carefully continuum normalized.
The majority of spectra were collected for a large ESO observing program on HARPS,
which aims at hunting for planets around solar twins \citep{Bedell2015, Melendez2017, Melendez2015},
complemented with HARPS spectra from other programs extracted from the online ESO Science Archive Facility.
The spectral signal-to-noise ratio SNR, measured at 6000\,{\AA},
has an average value of about 800 per pixel, varying from 300 up to 1800 per pixel.
The solar reference spectrum is a combined spectrum of the reflected light by the asteroid Vesta.
Its flux is continuum normalized in the same way as the spectra of the solar twins,
and it has SNR\,=\,1300 per pixel at 6000\,{\AA}.
Further details about the analysed HARPS spectra are given in \citet{Spina2018}.
Table~\ref{tab_sample} shows the main parameters of sample solar twins,
covering 58 stars as described at the end of the next section.

%
%
\begin{table*}
\centering
\caption{
Stellar parameters of the sample stars collected from previous published works
(covering 58 solar twins):
photospheric parameters and isochrone age \citep{Spina2018},
macro-turbulence and rotation velocities \citep{dosSantos2016}, and
elemental abundances \citep{Bedell2018}.
Full table online.
}
\label{tab_sample}
\resizebox{\linewidth}{!}{%
\begin{tabular}{rrrrrrrrrrr}
\hline
Star ID     & $T_{\rm eff}$ & log\,$g$     & [Fe/H]           & $\xi$         & $V_{\rm macro}$ & $V$.sin($i$) & age  & [Si/H] & [Nd/H]           & [Eu/H] \\
            &        (K) &                 & (dex)            & (km\,s$^{-1}$)& (km\,s$^{-1})$  &(km\,s$^{-1}$)& (Gyr)&  (dex) &  (dex)           &  (dex) \\
\hline
Sun         & 5777       & 4.440           &  0.000           & 1.00          & 3.20 & 2.04 & 4.56          &  0.000           &  0.000           &  0.000           \\
HIP\,003203 & 5868$\pm$9 & 4.540$\pm$0.016 & -0.050$\pm$0.007 & 1.16$\pm$0.02 & 3.27 & 3.82 & 0.50$\pm$0.30 & -0.112$\pm$0.009 &  0.160$\pm$0.017 &  0.057$\pm$0.014 \\
HIP\,004909 & 5861$\pm$7 & 4.500$\pm$0.016 &  0.048$\pm$0.006 & 1.11$\pm$0.01 & 3.33 & 4.01 & 0.60$\pm$0.40 & -0.047$\pm$0.012 &  0.219$\pm$0.012 &  0.099$\pm$0.013 \\
HIP\,007585 & 5822$\pm$3 & 4.445$\pm$0.008 &  0.083$\pm$0.003 & 1.01$\pm$0.01 & 3.37 & 1.90 & 3.50$\pm$0.40 &  0.058$\pm$0.008 &  0.138$\pm$0.009 &  0.124$\pm$0.015 \\
HIP\,008507 & 5717$\pm$3 & 4.460$\pm$0.011 & -0.099$\pm$0.003 & 0.96$\pm$0.01 & 2.88 & 0.77 & 4.90$\pm$0.45 & -0.127$\pm$0.028 &  0.044$\pm$0.012 & -0.019$\pm$0.022 \\
--          & --         & --              &  --              & --            & --   & --   &  --           &  --              &  --              &  --              \\
HIP\,118115 & 5798$\pm$4 & 4.275$\pm$0.011 & -0.036$\pm$0.003 & 1.10$\pm$0.01 & 3.55 & 0.89 & 8.00$\pm$0.30 & -0.029$\pm$0.012 &  0.023$\pm$0.006 & 0.069$\pm$0.007 \\
\hline
\end{tabular}%
}
\end{table*}

%
\section{Homogeneous thorium abundances}
\label{methodology}

\begin{figure*}
\includegraphics[width=75mm]{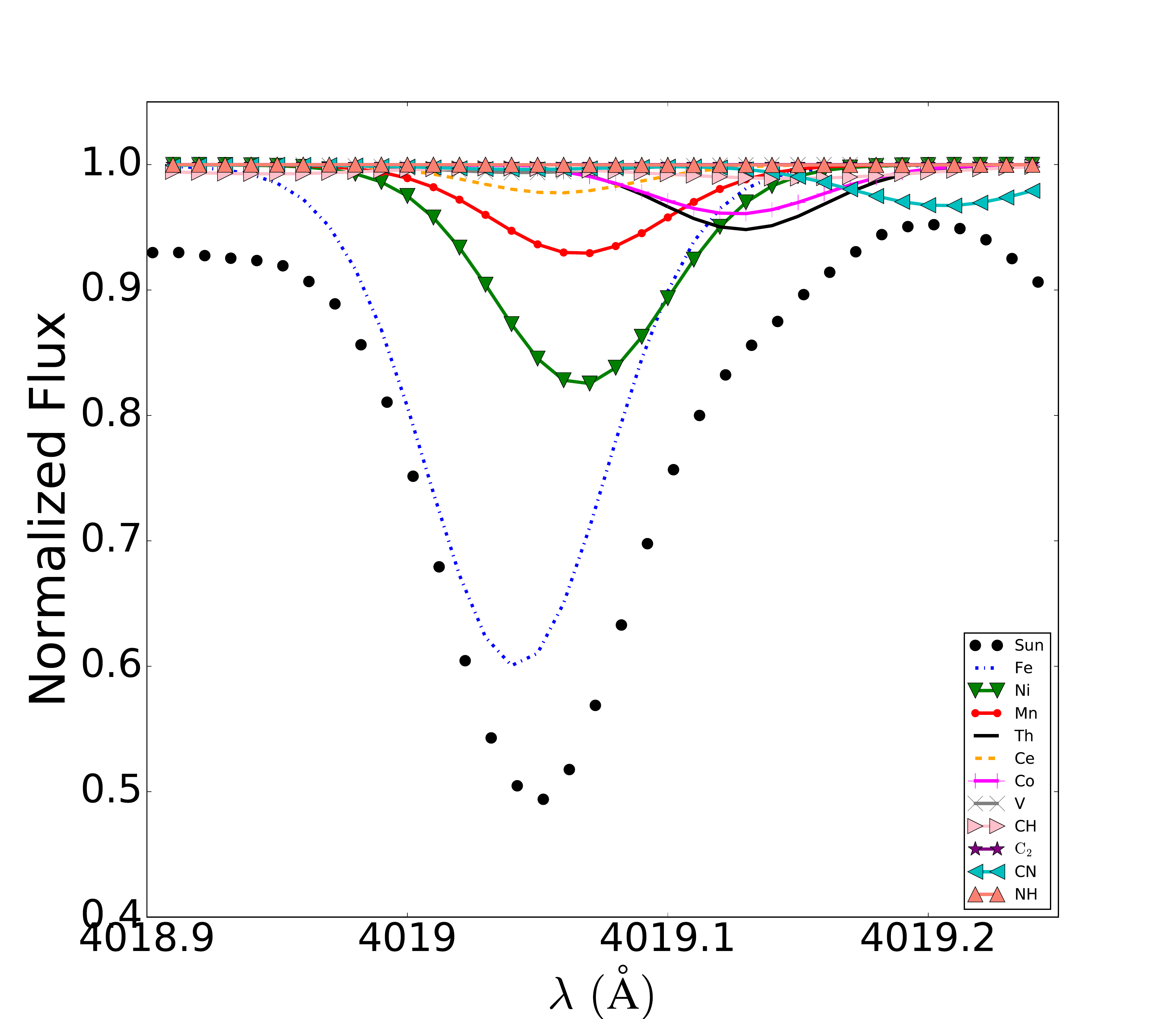}
\includegraphics[width=100mm]{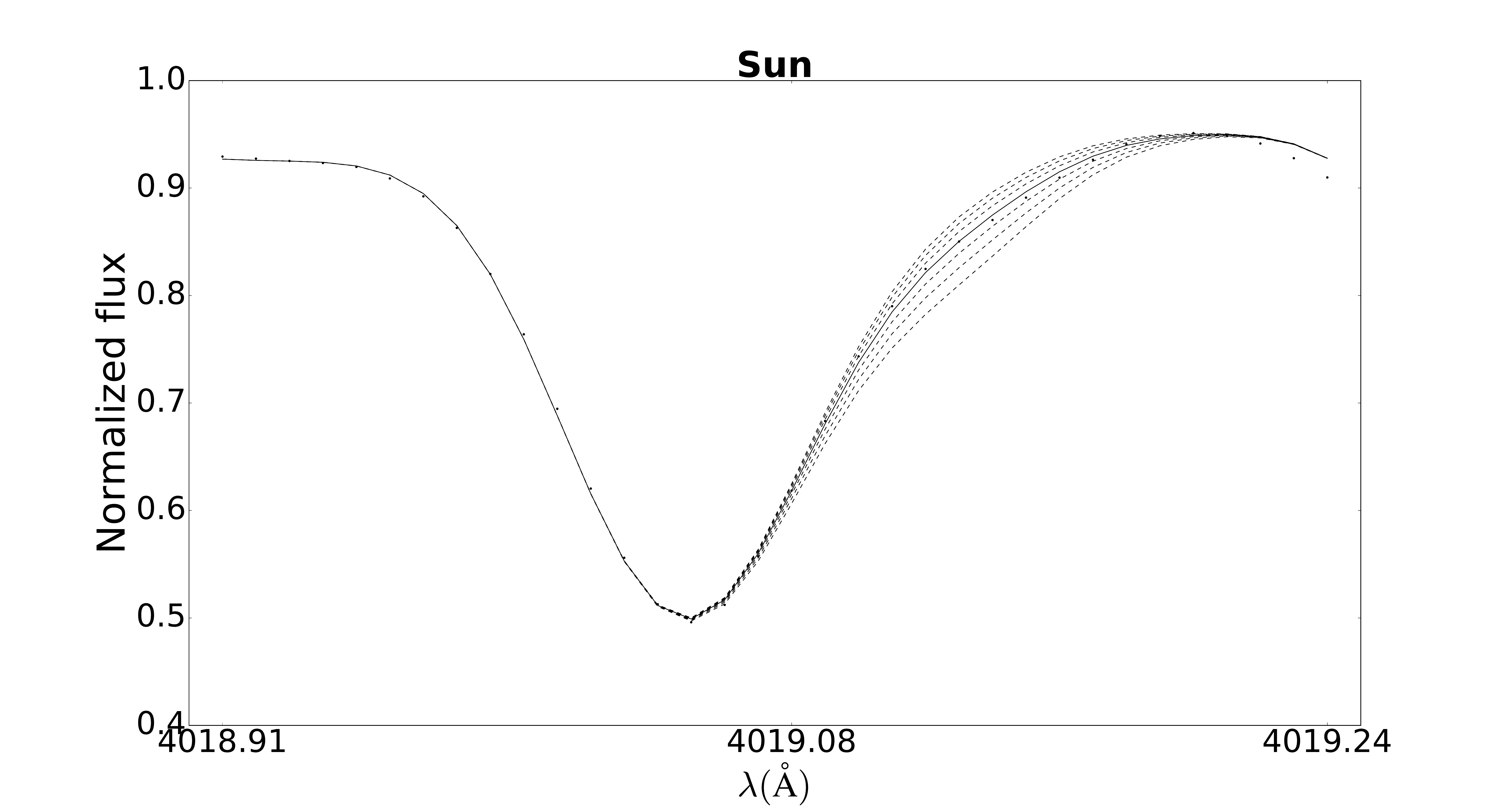}
\includegraphics[width=87mm]{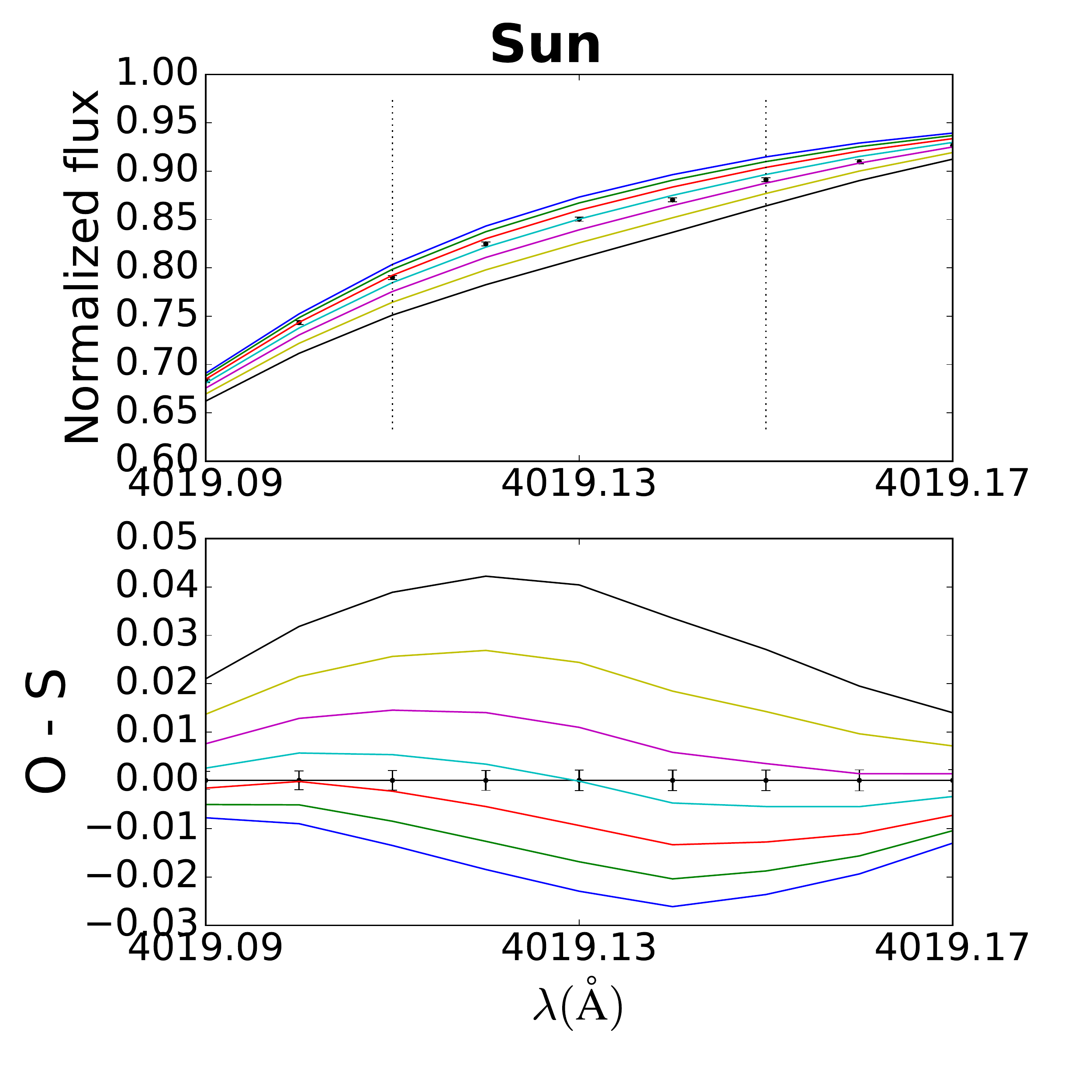}
\includegraphics[width=87mm]{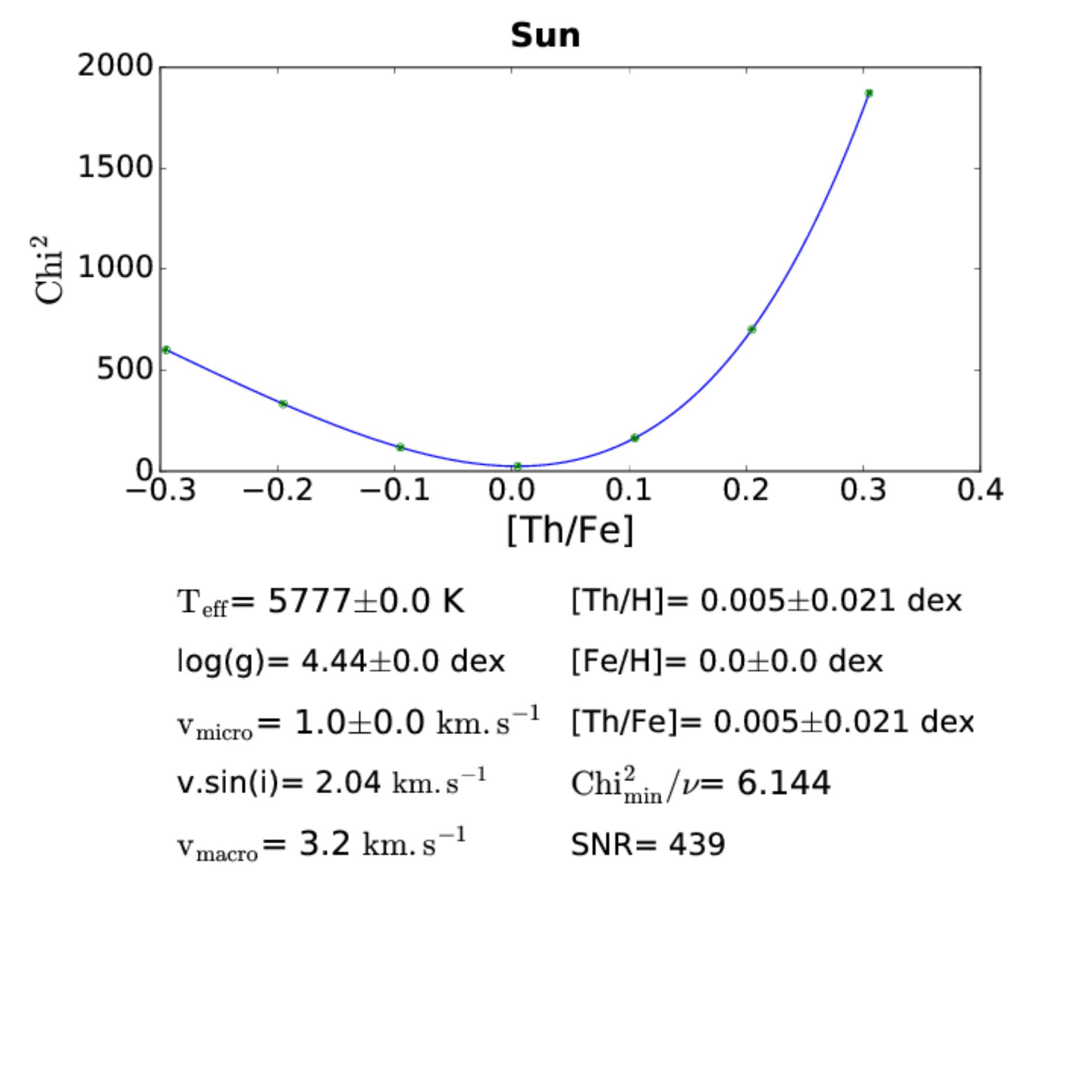}
\caption{
Spectral synthesis calibration of the Fe-Ni-Mn-Th-Co-CN-Ce-CH blend at 4019\,{\AA} to the solar spectrum
after the procedure of $gf$-values adjustment as explained in Sect.~\ref{methodology}:
multi-species atomic and molecular line contributions at the left top panel,
blended global profile reproduction at the right top panel,
comparisons between the observed spectrum and seven synthetic spectra at the left bottom panel,
and $\chi^{2}$ minimization graph at the right bottom panel.
The Th\,II 4019.1290-{\AA} line is present in the red wing.
}
\label{fig_blend_Sun_calib}
\end{figure*}

The abundance of Th has been measured from a complex blend at 4019\,{\AA},
which has been quite employed for this purpose, both in the study of Galactic disc stars
(e.g., \citet{Unterborn2015}, \citet{delPeloso2005b}, \citet{Morell1992})
and Galactic halo stars (e.g., \citet{Roederer2009}, \citet{JohnsonBolte2001}, \citet{Cowan1997}).
The analysis of this multi-species blend requires
a careful spectral synthesis at very high-resolution and high SNR.
The Th\,II 4019.1290-{\AA} line together with
the Co\,I 4019.1260-{\AA}, CH 4019.1390-{\AA} and CN 4019.2060-{\AA} lines
make the blend asymmetric at the red wing.
The very weak CH line is from the (3,2) vibrational band of the A-X electronic system ($^{12}$C$^{1}$H),
and the relatively weak CN line is from the (4,5) vibrational band of the B-X electronic system ($^{12}$C$^{14}$N).
In the case of solar-type stars,
the major atomic contributor of this blend is the Fe\,I 4019.0420-{\AA} line
followed by the Ni\,I 4019.0667-{\AA} and Mn\,I 4019.0658-{\AA} lines.
A minor contribution comes from the Ce\,II 4019.0569-{\AA} line.
We refer the reader to Fig.~\ref{fig_blend_Sun_calib} to see the different contributions
to the blend, based on our calibration through spectral synthesis of the solar spectrum, as detailed in this section.

There is no mensurable line of uranium in the HARPS spectral coverage.
However there is another Th\,II line in the HARPS range (at 4086.5210\,{\AA}),
but it is 4.5 times weaker than the Th\,II 4019.1290-{\AA} line.
Uranium is another $r$-process radioactive element that also contributes
for the energy budget of terrestrial planets (mantle convection/thickness).


%

The spectral synthesis has been performed in the region $\lambda\lambda$4000--4030\,{\AA}
using the 2014 version of the MOOG code \citep{Sneden1973}
that solves the photospheric radiative transfer with line formation
in FGK stars under the LTE (local thermodynamic equilibrium) approximation.
The spectral synthesis based on MOOG has been handled
through our own automated procedure developed under the Python language (version 2.7),
which covers several steps
from the solar $gf$ calibration to the derivation of the Th abundance with its uncertainty,
as explained in this section.
We have also adopted the same stellar parameters and model atmospheres
as our previous analyses of the sample stars \citep{Spina2018, Bedell2018} to be self-consistent with them,
i.e. interpolated models in the grid of \citet{CastelliKurucz2004} (ATLAS9) assuming solar abundance ratios.
Throughout our differential analysis we have also adopted the standard parameters for the Sun:
$T_{\rm eff}$\,=\,5777\,K, log\,$g$\,=\,4.44, [Fe/H]\,=\,0.00\,dex
and $\xi$\,=\,1.00\,km\,s$^{-1}$
(e.g. \citep{Cox2000}).
The solar chemical composition by \citet{Asplund2009},
and by \citet{Grevesse2015} specifically for the heavy elements Cu to Th,
have been adopted as reference,
such that the adopted Sun's Th abundance is log$\epsilon$(Th)=0.03\,dex.
The wavelength step of synthetic spectra has been fixed to 0.01\,{\AA}.

The atomic line list has been compiled from the VALD database \citep{Kupka2000},
and we have added the following molecular lines from the Kurucz database \citep{Kurucz2017}:
$^{12}$C$^{1}$H and $^{13}$C$^{1}$H lines of the A-X system,
$^{12}$C$^{1}$H and $^{13}$C$^{1}$H lines of the B-X system,
$^{12}$C$^{12}$C lines of the d-a system,
$^{12}$C$^{14}$N lines of the B-X system, and  
$^{14}$N$^{1}$H lines of the A-X system.
In the case of solar-like stars,
only lines of the A-X system of $^{12}$CH, B-X system of $^{12}$CH,
d-a system of $^{12}$C$^{12}$C and B-X system of $^{12}$C$^{14}$N,
actually produce detectable absorptions in the studied region at our SNR ($\sim$ 800).
The other molecular lines (the main isotopes of carbon and nitrogen)
are included in the line list to make more realistic the reproduction of the stellar continuum,
because they are supposed to be within the spectral noise for SNR\,$\leq$\,1000 ($^{13}$CH B-X system),
or even 10,000 ($^{13}$CH A-X system, $^{14}$NH A-X system).
We have fixed solar isotopic ratios for $^{12}$C/$^{13}$C and $^{14}$N/$^{15}$N
that are, respectively, 89.4 and 435 \citep{Asplund2009}.
We have also confirmed that the lines of CH A-X and CN B-X
in the investigated 30\,{\AA}-wide region are all insensitive to variation of C and N isotopic ratios.
There are very few lines of CH B-X that are somehow sensitive to the $^{12}$C/$^{13}$C ratio,
but none affects the analysed Fe-Ni-Mn-Th-Co-CN-Ce-CH blended feature at $\lambda\lambda$4018.9--4019.2\,{\AA}.
The adopted dissociative energies of CH, C$_{2}$, CN and NH are those by \citet{BarklemCollet2006}.
We tested somewhat different values of the dissociation constants,
and found no changes in the results within about 0.01\,dex.
Table~\ref{tab_linelist} presents the list of atomic and molecular lines
covering 2\,{\AA} around the Th\,II feature, whose $gf$ values
have been calibrated to the cited solar spectrum, as explained below.

%
%
\begin{table}
\centering
\caption{
Line list after the $gf$ calibration to the solar spectrum:
wavelength,species code, excitation potential of transition inferior level, log\,$gf$ and species identification.
The list covers 2\,{\AA} around the Fe-Ni-Mn-Th-Co-CN-Ce-CH blend at 4019\,{\AA}.
The species code is the MOOG standard notation,
i.e. atomic number(s) before the decimal point (listed in crescent order for molecules)
followed by the ionization level immediately after the decimal point (0: neutral, 1: first ionized, and so on)
and also by mass numbers for molecules in the case of isotopic species discriminated (listed in crescent order).
Full table online.
}
\label{tab_linelist}
\begin{tabular}{rlrrl}
\hline
wavelength & species code & $\chi_{e}$ & $gf$     & species \\
   ({\AA}) &              &       (eV) &          &        \\
\hline
4018.5080  & 106.00113   &     1.391 & 0.276E-05 & CH \\
4018.5090  & 607.0       &     3.754 & 0.154E+00 & CN \\
4018.5360  & 607.0       &     3.193 & 0.424E-04 & CN \\
4018.5620  & 107.0       &     2.440 & 0.433E-02 & NH \\
4018.5703  & 25.0        &     5.087 & 0.195E+00 & Mn\,I \\
---        & ---         &     ---   & ---       & --- \\
4020.4940  & 106.00113   &     0.595 & 0.162E-04 & CH \\
\hline
\end{tabular}
\end{table}

The first step for the differential chemical analysis relatively to the Sun using spectral synthesis
is the calibration of $gf$-values for reproducing the line profiles,
adopting a solar spectrum obtained with the same instrument and resolution.
We have performed the calibration of $gf$-values of all lines in the region $\lambda\lambda$4000--4030\,{\AA}
through an automatic procedure under a line-by-line basis, using the flux at the core of each individual line.
Specifically for a very narrow region around the 4019-{\AA} blend (about 1{\AA} wide),
we have refined the $gf$ calibration focusing on the contribution lines of the blend itself
paying a special attention to the continuum adjustment between the observed and synthetic spectra.
Due to the very close proximity between the cores
of the Th\,II and Co\,I lines under the HARPS spectral sampling
(respectively 4019.1290 and 4019.1260\,{\AA}),
we have fixed the $gf$-values for both lines, adopting accurate known experimental values
(0.592$\pm$0.018 for the Th\,II line by \citet{Nilsson2002}
and 0.537$\pm$0.049E-02 for the Co\,I line by \citet{Lawler1990}).
Whilst the $gf$-value of the Co\,I line has been actually fixed,
the $gf$-value of the Th\,II line has been slightly refined
to reproduce the spectral blend against the solar spectrum,
ensuring a confident differential analysis to the Sun
(the $gf$ fine tuning of the Th\,II line
has been just 1.9\,per\,cent greater than the experimental measurement
that has a relative error of 3\,per\,cent).
We have got [Th/H]\,=\,[Th/Fe]\,=\,0.005\,dex over this calibration step relatively to the Sun.
Figure~\ref{fig_blend_Sun_calib} illustrates
the spectral synthesis calibration of the Fe-Ni-Mn-Th-Co-CN-Ce-CH blend at 4019\,{\AA} to the solar spectrum.
We have been also able to confirm the relative intensity between the Th\,II and Co\,I lines
as semi-theoretically predicted by \citet{Grevesse2015}.
The ratio EW(Th\,II\,4019.1290-{\AA}\,line)/EW(Co\,I\,4019.1260-{\AA}\,line) agrees within 1 sigma deviation
taking into account the 20\,per\,cent estimated uncertainty for the EW(Th\,II\,4019.1290-{\AA}\,line) made by them.
To finely reproduce the line broadening,
we have adopted the line-of-sight rotational velocity $V$.sin($i$)
and macro-turbulence velocity $V_{\rm macro}$ that were respectively measured and computed by
\citet{dosSantos2016} for our sample solar twins
(the later one parametrically as a function of $T_{\rm eff}$ and log\,$g$).
A limb darkening linear law for solar-type stars has been employed
(limb darkening parameter $u$\,=\,0.60 adopted,
to be consistent with \citet{dosSantos2016}).
Based on our calibration to the HARPS solar spectrum
we estimate that the CH 4019.1390-{\AA} and CN 4019.2060-{\AA} lines are
81 and 28\,per\,cent weaker than the Th\,II line, respectively.
We have computed their EW in each individual synthetic spectrum.
Note that the cores of CH and CN lines are, respectively, 0.010 and 0.077\,{\AA} far away
from the core of Th\,II line (see Fig.~\ref{fig_blend_Sun_calib}).

%
\begin{figure*}
\includegraphics[width=115mm]{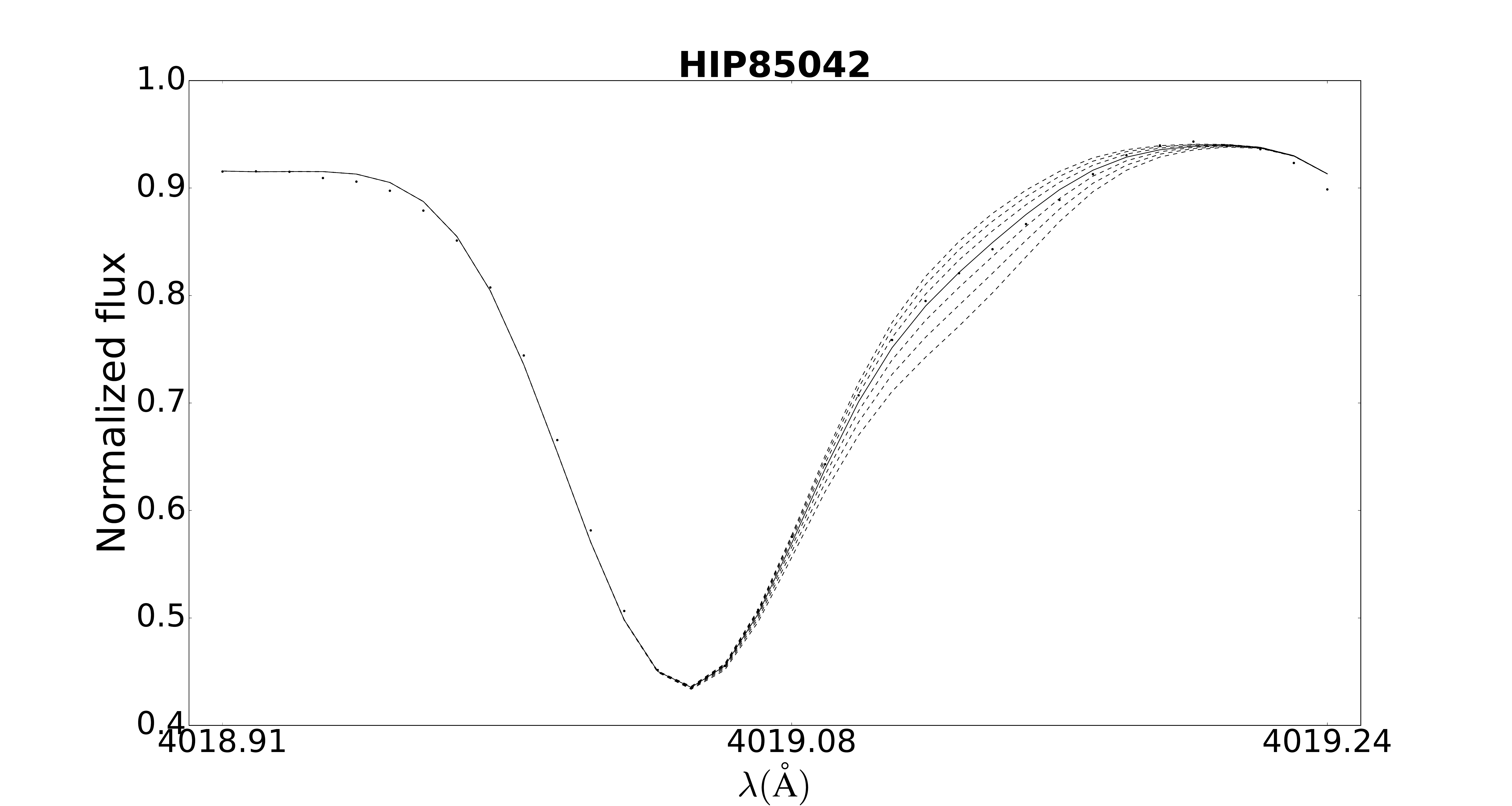}
\includegraphics[width=60mm]{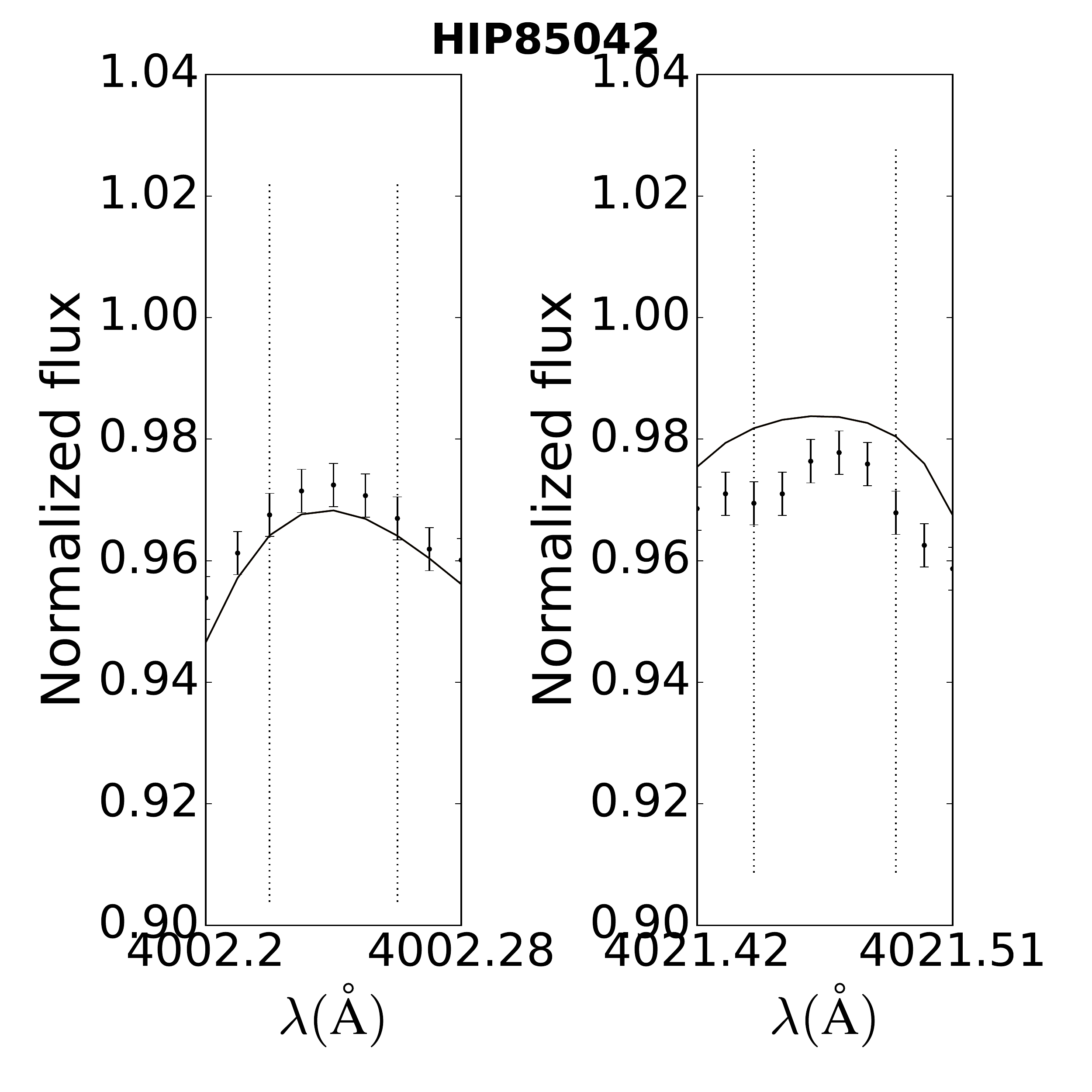}
\includegraphics[width=87mm]{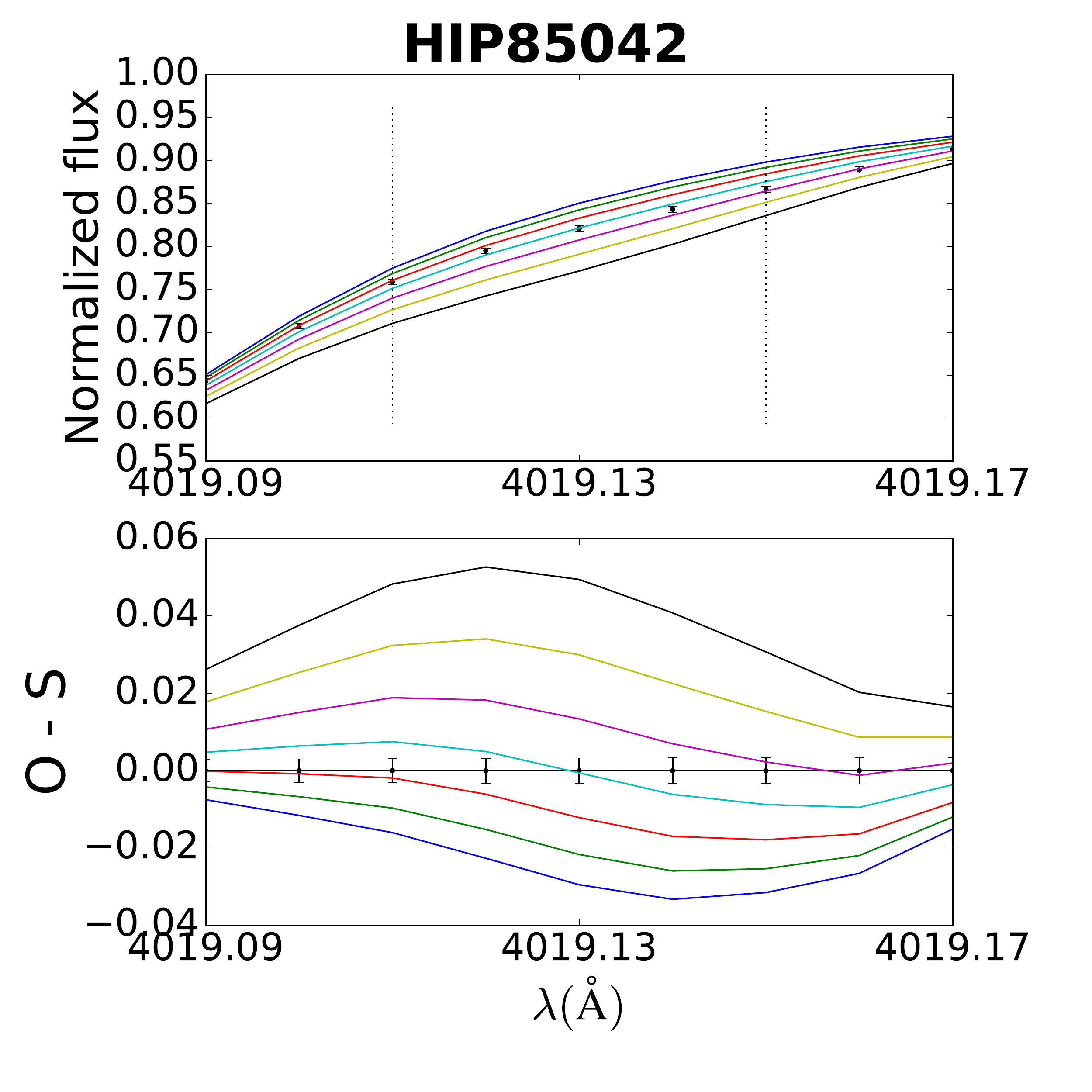}
\includegraphics[width=87mm]{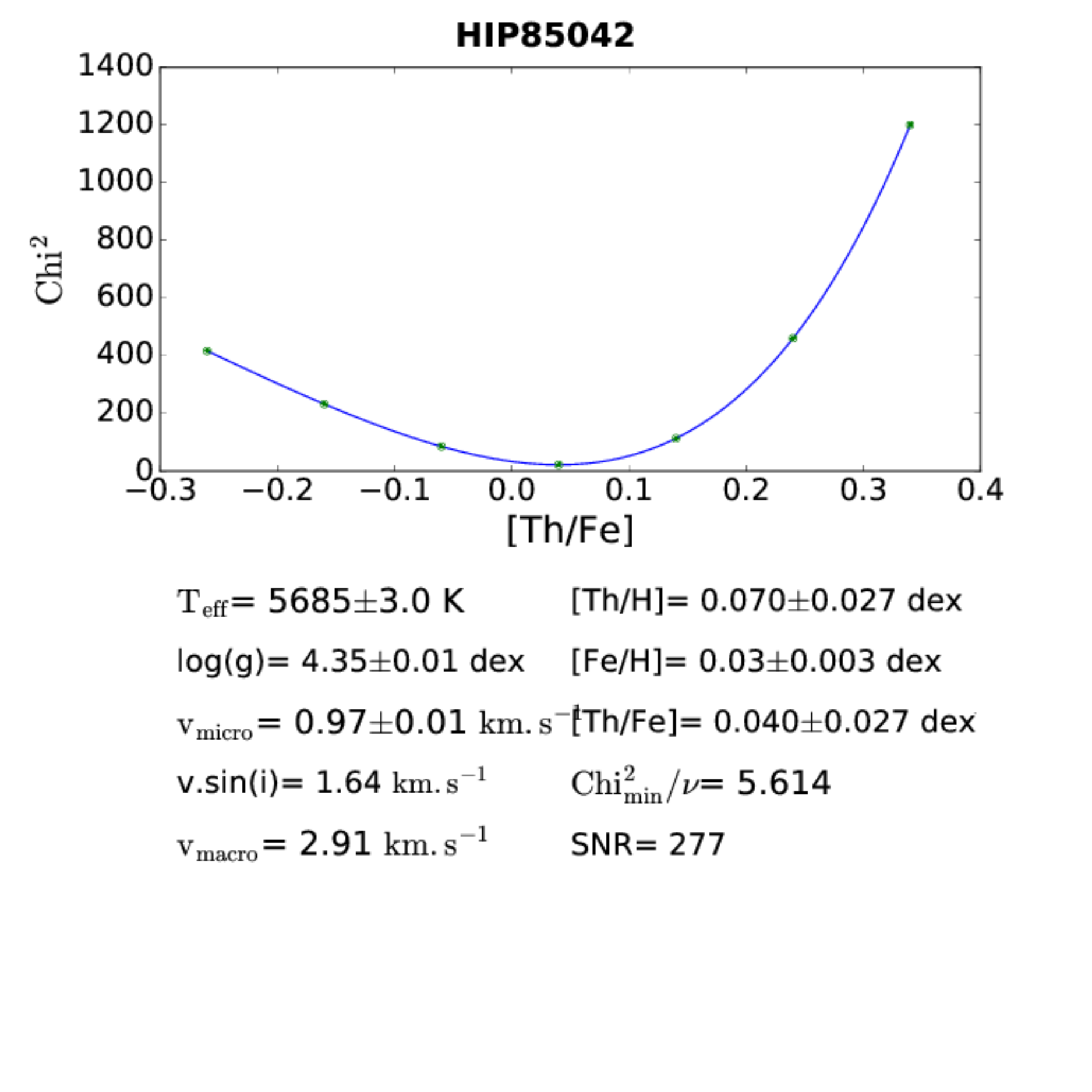}
\caption{
HIP\,85042: three spectral synthesis diagnostic plots (global profile, continuum windows and spectral comparisons)
and $\chi^{2}$ minimization graph (right bottom panel)
for the Fe-Ni-Mn-Th-Co-CN-Ce-CH blend at 4019\,{\AA},
whose red wing is sensitive to the Th abundance variation (Th\,II 4019.1290\,{\AA}-line).
Seven theoretical spectra are computed assuming a uniform step of 0.10\,dex in [Th/Fe].
A polynomial fit is applied for $\chi^{2}$ as a function of [Th/Fe]
to search for the minimum value and the resulting Th abundance ratio.
}
\label{fig_synthesis}
\end{figure*}

An automatic procedure based on the $\chi^{2}$ minimization has been used
for performing the spectral synthesis fit in order to extract the Th abundance from the blended feature
($\chi^{2}$\,=\,$\sum_{i{\,=\,}1,5}$ $(O_{i} - S_{i})^{2}$/$\sigma(O_{i})^{2}$, where 
$O_{i}$ is the flux of the observed spectrum,
$S_{i}$ is the flux of the synthetic spectrum,
$\sigma(O_{i})$ is the error in the observed flux,
and $i$ represents the wavelength point).
The wavelength window to compute $\chi^{2}$ is centered
at the core of Th\,II line, as the spectral sampling (i.e. 4019.13\,{\AA}),
and it covers a region in which the line absorption flux is stronger than or comparable to the other lines
taking the solar spectrum as reference (see Fig.~\ref{fig_blend_Sun_calib}).
The $\chi^{2}$ window is $\lambda\lambda$4019.11--4019.15\,{\AA}, ranging over five pixels.
The error in observed flux is estimated as a function of the continuum SNR,
such that $\sigma(O_{i})$ = $O_{i}$/SNR$_{\rm continuum}$.
The continuum SNR in the blend region has been measured as a simple average over five small continuum windows
(centered at 4002.24\,{\AA}, 4009.35\,{\AA}, 4019.67\,{\AA}, 4020.67\,{\AA} and 4021.47\,{\AA}).
The local SNR ranges from 147 (HIP\,8507) up to 993 (HIP\,54582),
having an average value of 379 ($\sigma$(SNR)\,=\,152).
For the Sun's spectrum, we have SNR\,=\,439.
The normalization of flux scale is fine tuned for the absorption line blend
by applying a multiplicative correction factor taking as reference
the continuum points at 4002.24\,{\AA} and/or 4021.47\,{\AA}.

The $\chi^{2}$ procedure is particularly useful to directly estimate the error in Th abundance
due to the spectral synthesis fitting itself,
such that the error in [Th/Fe] is derived when $\chi^{2}$ increases in $\nu$ freedom degree unities
from the minimum $\chi^{2}$ value along the polynomial fitting curve $\chi^{2}$ vs. [Th/Fe].
In this case, the freedom degree $\nu$ is equal to 4, 
because the $\chi^{2}$ window covers 5 pixels and there is a single free parameter that is the Th abundance itself
(i.e. $\nu$\,=\,n$_{\rm pixels}$ - n$_{\rm parameters}$).
The thorium abundance uncertainties due to the propagation of the errors of photospheric parameters
have been added in quadrature to the error of the spectral synthesis.
Figure~\ref{fig_synthesis} illustrates the spectral synthesis procedure of the Th\,II line in the case of HIP\,85042.

We have also verified the influence of errors in the abundances of carbon and cobalt in the Th abundance error
due to the close proximity of those lines of Co\,I, CH and CN to the core of the Th\,II line.
Taking the case of HIP\,85042 as representative for the sample stars,
the uncertainties in [Co/H] and [C/H] \citep{Bedell2018}
make changes in [Th/Fe] correspondent respectively to 0.003 and 0.002\,dex,
which are many times smaller than the other involved errors.
Therefore the impact of Co and C abundance errors in the Th abundance analysis is negligible, and  
the final error in [Th/Fe] is basically fixed by the spectral synthesis error itself.
The average error in [Th/Fe] is 0.025\,dex considering 53 stars in total,
after removing objects whose derived Th abundances are not acceptable, as justified  
in the following paragraph. Finally, the error in [Th/H] is computed in quadrature
taking into account the error in [Fe/H], and it coincides with that one in [Th/Fe].

The final step for measuring the Th abundance
is to check the quality of spectral synthesis of the whole analysed absorption blend
on a star-by-star basis (sample of 67 stars).
First of all, the fit is very unsatisfactory in the cases of HIP\,65708 and HIP\,83276
(the later certainly due to an unidentified problem in the data reduction).
Results are not acceptable due to the low spectral SNR at the studied 30\,{\AA}-wide region
for HIP\,6407, HIP\,81746 and HIP\,89650
(roughly for SNR\,$\lesssim$\,130 and/or $\sigma$[Th/Fe]\,$\gtrsim$\,0.070\,dex).
The blend has not been reproduced well
for others 4 stars:
HIP\,10303, HIP\,14501, HIP\,30037, and HIP\,115577
(two of them are alpha-rich old solar twins
and there is nothing special related to the two others).
Three out of these 9 eliminated stars (HIP\,14501, HIP\,65708, and HIP\,115577)
were also excluded from the fits made in the chemical evolution study of \citet{Bedell2018}
because they were supposed to belong to the thick disc
(ages above 8\,Gyr and a visible enhancement in $\alpha$-elements).

For fifteen cases
(HIP\,3203,
HIP\,7585,
HIP\,18844,
HIP\,22263,
HIP\,25670,
HIP\,28066,
HIP\,30476,
HIP\,34511,
HIP\,40133,
HIP\,42333,
HIP\,73241,
HIP\,74432,
HIP\,79672,
HIP\,101905 and
HIP\,104045),
we have improved the automated spectral synthesis fit by slightly changing the continuum level.
We have also revised the spectral synthesis fit of the Th\,II blended feature
for a few very young stars of our sample (4 stars in total)
by modifying the rotational velocity in order to better globally reproduce the feature
(decrease by about 8-20\,per\,cent in $V$.sin($i$)).
They are
HIP\,3203,
HIP\,4909,
HIP\,38072, 
and HIP\,101905,
having respectively a decrease of 8, 10, 20 and 13\,per\,cent in $V$.sin($i$).
Despite of few cases, for which the rotational broadening had to be modified,
we can state that the Th abundance is more sensitive to the continuum level than the rotational broadening.

We have investigated for several cases the impact of changing the abundance of Fe
in order to get a better spectral synthesis fit
since the blended feature is dominated by a Fe line, specially at the blue wing.
We conclude that the derived Th abundance is just slightly modified within its error,
even when the log\,$\epsilon$(Fe) increases or decreases up to 5-10 times its error.
The same conclusion is reached by changing the abundances of Ni, Mn and/or Ce,
whose lines shape the blended red wing.
Consequently, their derived Th abundances are not updated
upon the effects of changing the Fe, Ni and Mn individual abundances.

In order to be consistent with the work of \citet{Bedell2018},
we have decided to exclude from our analysis
HIP\,28066,
HIP\,30476,
HIP\,73241,
HIP\,74432
because they are $\alpha$-rich old solar twins,
as well as HIP\,64150 because it exhibits anomalous $s$-process element abundances
likely due to a past contamination from a close binary companion
\citep{Bedell2018, Spina2018, dosSantos2017}.

%
%

In total we have measured and make available Th abundance for 58 solar twins.
However, just 53 thin disc solar twins are adopted in our analysis of correlations
of the Th abundance as function of [Fe/H] and isochrone stellar age
(1 chemically anomalous in $s$-elements and 4 $\alpha$-rich old stars are excluded).
The Sun is not included in the analysis too.
Table~\ref{tab_th_results} shows the derived Th abundances in 58 solar twin stars.

%
%
\begin{table}
\centering
\caption{
Thorium abundance measured in this work relatively to the Sun for 58 solar twins (Sun included as well).
Full table online.
}
\label{tab_th_results}
\resizebox{\linewidth}{!}{%
\begin{tabular}{rrrrrrr}
\hline 
Star ID     & [Th/H] & error & [Th/Fe] & error & SNR & $\chi^{2}_{min}/\nu$ \\
            &   (dex) & (dex) &  (dex) & (dex) &     &                \\
\hline 
Sun         & +0.005  & 0.021 & +0.005  & 0.021 & 439 &  6.144 \\
HIP\,003203 & +0.120  & 0.034 & +0.170  & 0.033 & 319 &  0.041 \\
HIP\,004909 & +0.180  & 0.023 & +0.132  & 0.022 & 534 & 10.978 \\
HIP\,006407 & +0.004  & 0.081 & +0.062  & 0.081 & 121 &  1.901 \\
HIP\,007585 & +0.193  & 0.018 & +0.110  & 0.018 & 454 &  3.796 \\
--          & --      & --    &  --     & --    & --  & --     \\
HIP\,118115 & +0.009  & 0.026 & +0.045  & 0.026 & 295 &  0.071 \\
\hline
\end{tabular}
}
\end{table}

%
\section{Thorium abundance ratios versus metallicity and stellar age}
\label{results}

We have analysed the derived Th abundances for 53 thin disc solar twins
as a function of both metallicity and isochrone age.
In this section we compile our results.

\begin{figure}
\includegraphics[width=\columnwidth]{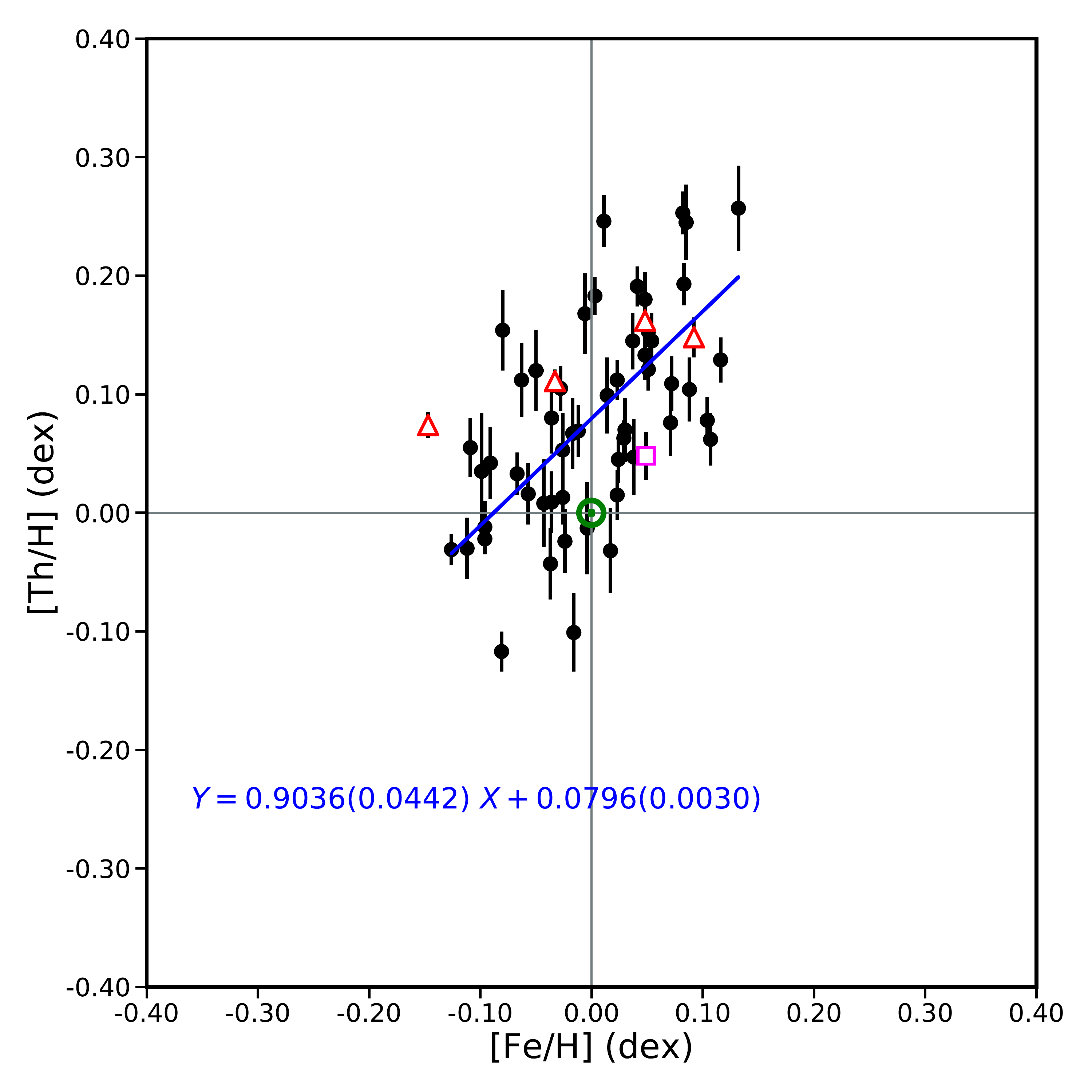}
\includegraphics[width=\columnwidth]{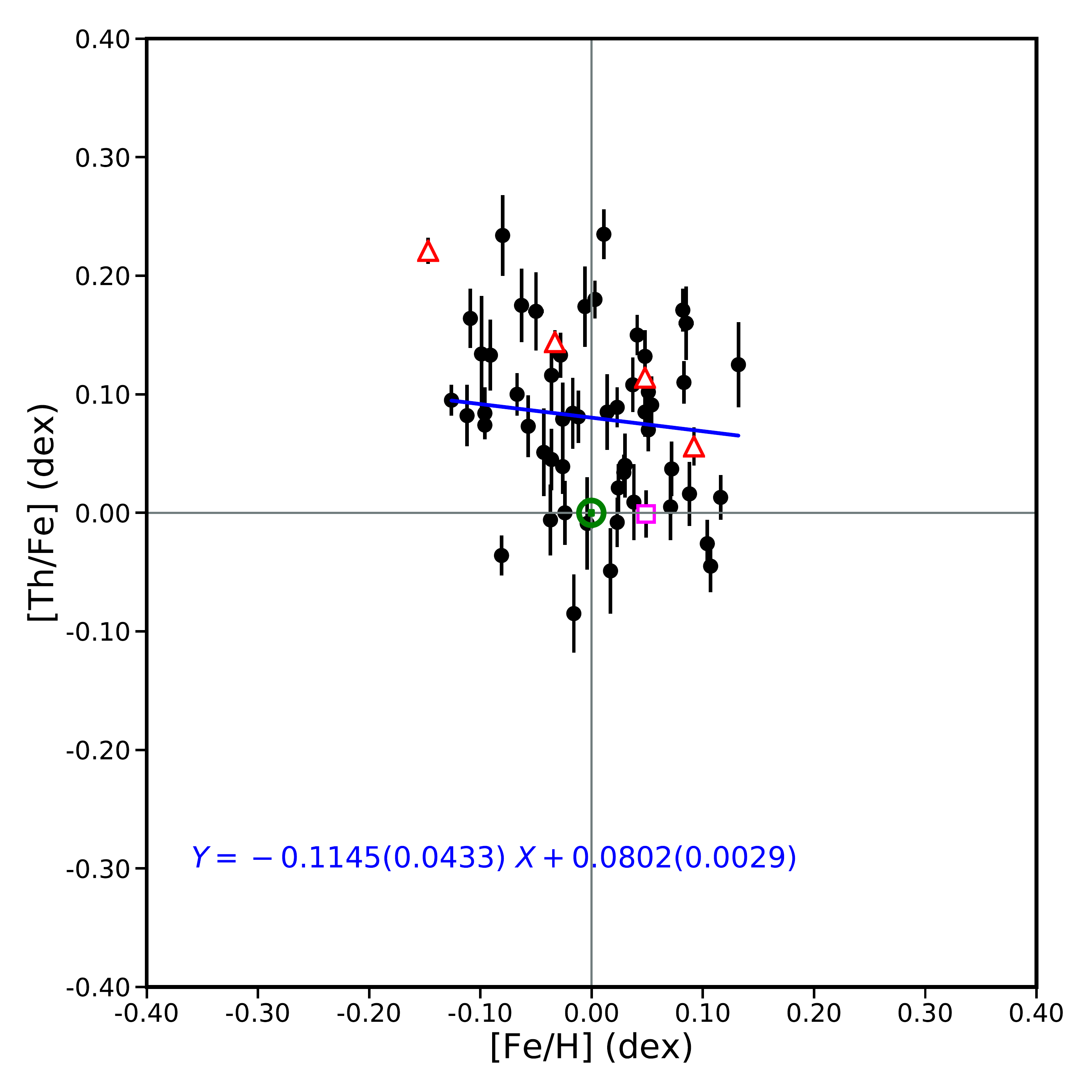}
\caption{
[Th/H]$_{\rm observed}$ and [Th/Fe]$_{\rm observed}$ as a function of [Fe/H]:
a statistically robust linear fit is shown in each plot (blue solid line).
The fitting equation with coefficients and their errors inside parenthesis are also displayed.
The fit corresponds to 53 thin disc solar twins only (black filled circles).
Data of excluded stars are plotted together for illustration purposes only
(red empty triangles: four $\alpha$-rich old stars,
and magenta empty square: the chemically anomalous in $s$-elements HIP\,64150).
Sun's data are also plotted as reference at the coordinates origin (green solar standard symbol),
although are consistent with the fits.
}
\label{fig_thh_thfe_obs_vs_feh}
\end{figure}

We have verified that the current (observed) Th abundance [Th/H] in solar-twin stars
follows the Fe abundance over the very restricted metallicity interval of them.
There is no (anti-)correlation between the observed values of [Th/Fe] and [Fe/H].
Figure~\ref{fig_thh_thfe_obs_vs_feh} shows the observed [Th/H] and [Th/Fe] as a function of [Fe/H],
in which linear fitting curves are drawn for illustration purposes only.
The fitting procedure is further described in this section.

\begin{figure*}
\includegraphics[width=80mm]{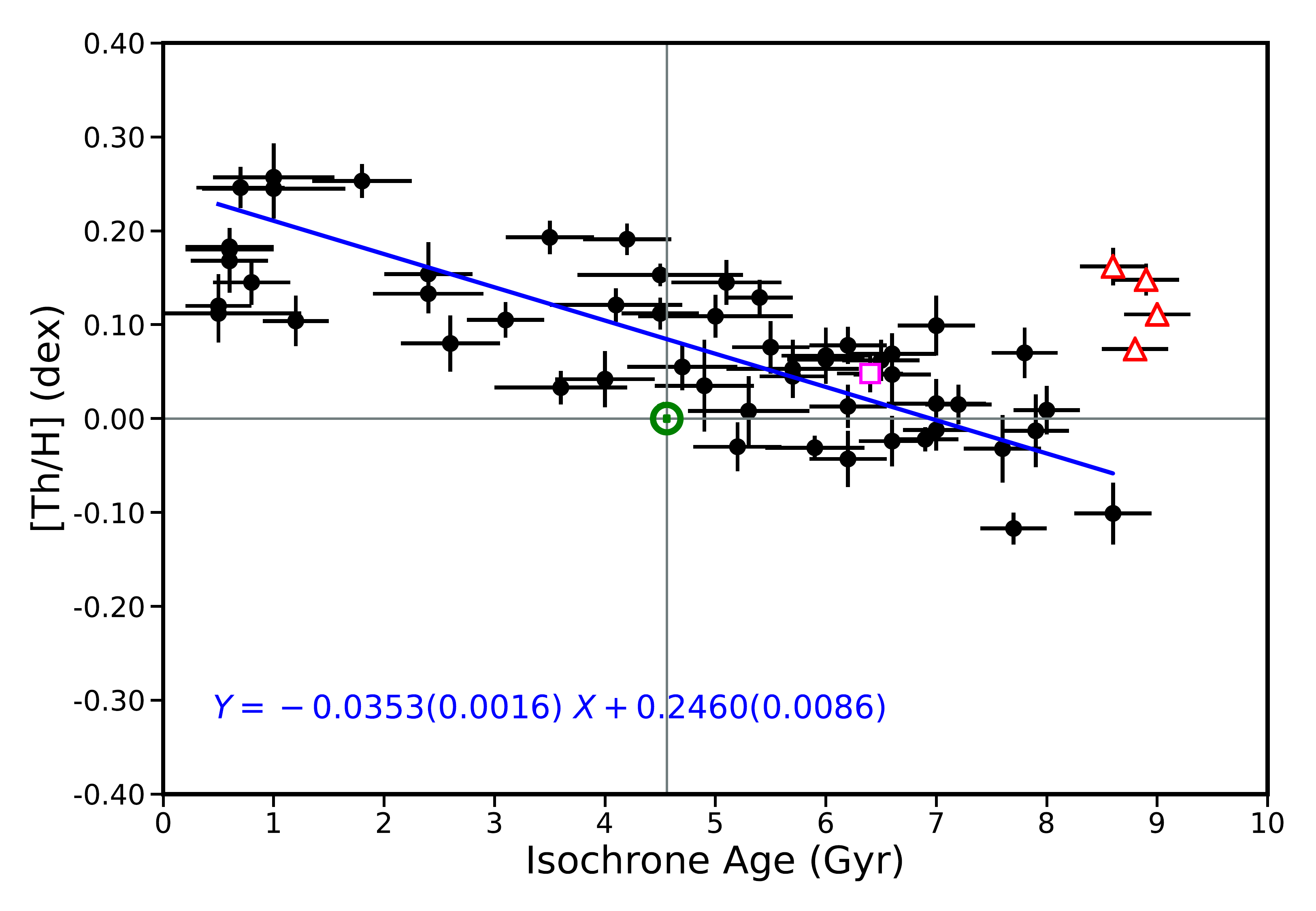}
\includegraphics[width=80mm]{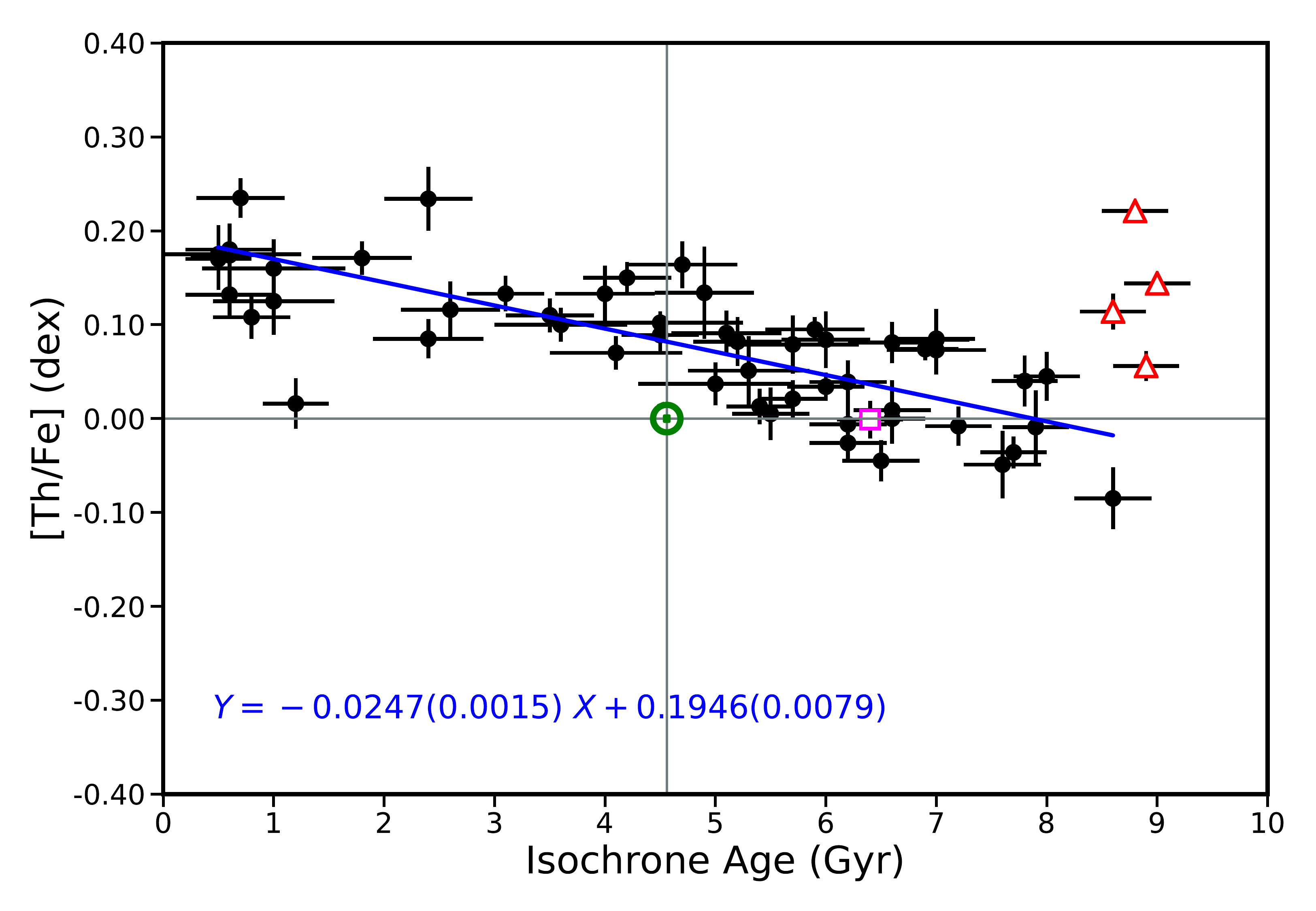}
\caption{
[Th/H]$_{\rm observed}$ and [Th/Fe]$_{\rm observed}$ as a function of the isochrone stellar age:
a statistically robust linear fit is shown in each plot (blue solid line).
The fitting equation with coefficients and their errors inside parenthesis are also displayed.
The fits correspond to 53 thin disc solar twins only (black filled circles).
Data of excluded stars are plotted together for illustration purposes only
(red empty triangles: four $\alpha$-rich old stars,
and magenta empty square: the chemically anomalous in $s$-elements HIP\,64150).
Sun's data are also plotted as reference (green solar standard symbol),
although are consistent with the fits.
}
\label{fig_thh_thfe_obs_vs_age}
\end{figure*}

The observed strong anti-correlations [Th/H]-Age and [Th/Fe]-Age (current [Th/H] and [Th/Fe] ratios)
are partially explained by the radioactive decay of Th, because after correcting the Th abundances by this effect,
the anti-correlation [Th/H]-Age keeps existing rather than being erased.
Figure~\ref{fig_thh_thfe_obs_vs_age}
shows the observed [Th/H] and [Th/Fe] as a function of the isochrone stellar age,
in which linear fitting curves are drawn for illustration purposes only.

%

%

%
\begin{figure}
\includegraphics[width=\columnwidth]{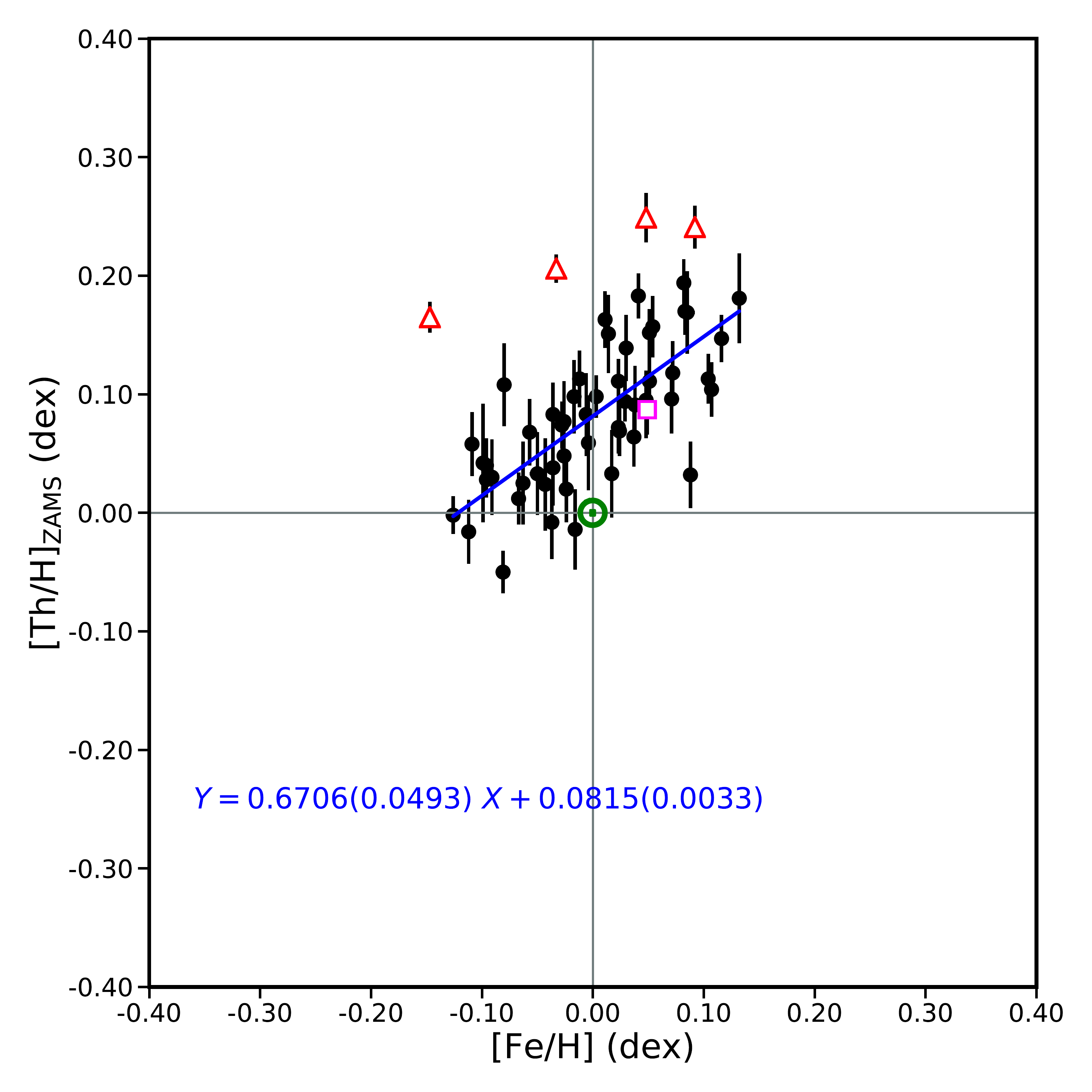}
\includegraphics[width=\columnwidth]{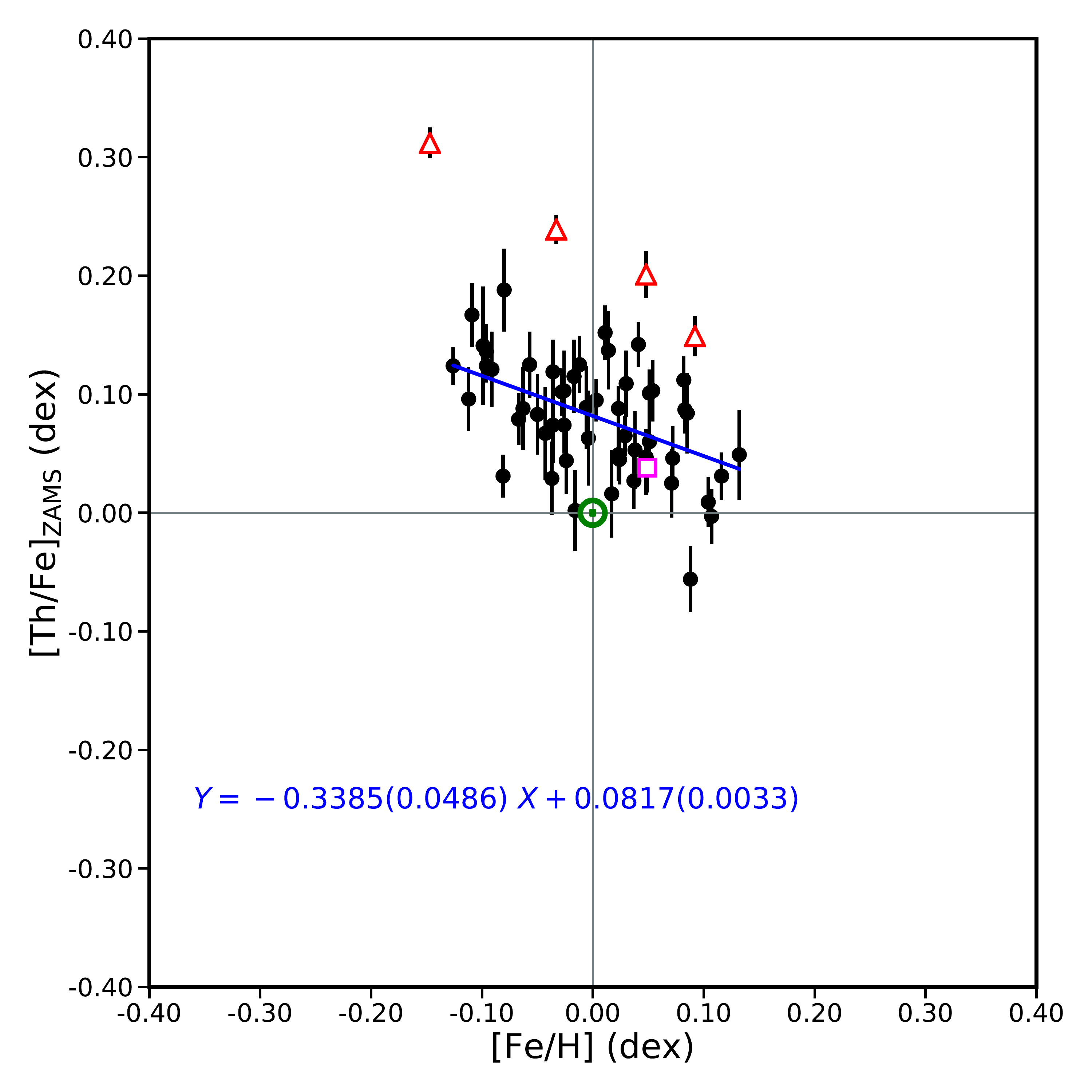}
\caption{
[Th/H]$_{\rm ZAMS}$ versus [Fe/H] and [Th/Fe]$_{\rm ZAMS}$ versus [Fe/H]:
a statistically robust linear fit is shown in each plot (blue solid line).
The fitting equation with coefficients and their errors inside parenthesis are also displayed.
The fit corresponds to 53 thin disc solar twins only (black filled circles).
Data of excluded stars are plotted together for illustration purposes only
(red empty triangles: four $\alpha$-rich old stars,
and magenta empty square: the chemically anomalous in $s$-elements HIP\,64150).
Sun's data are also plotted as reference at the coordinates origin (green solar standard symbol),
although are consistent with the fits.
}
\label{fig_thh_thfe_zams_vs_feh}
\end{figure}

%
\begin{figure*}
\includegraphics[width=80mm]{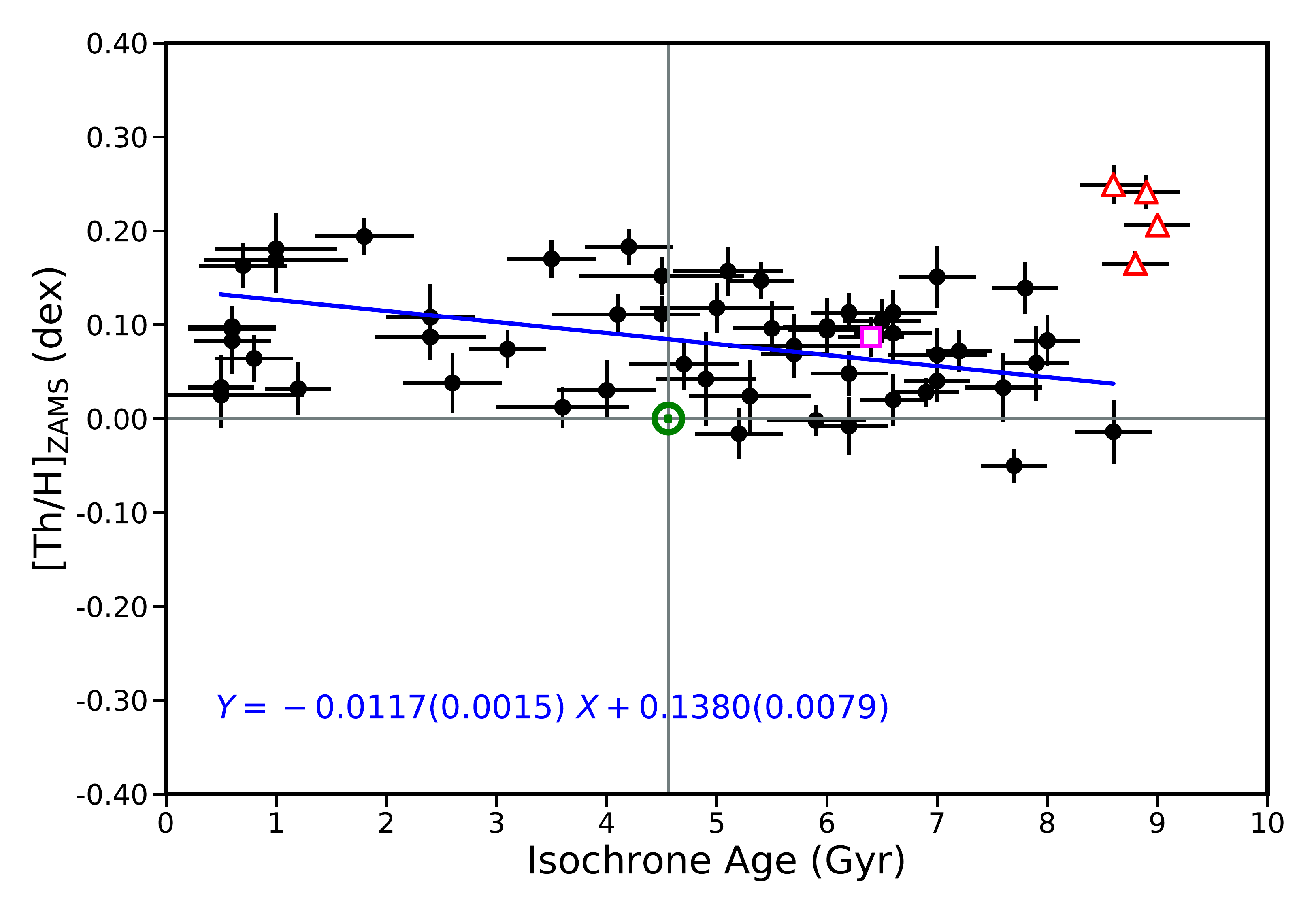}
\includegraphics[width=80mm]{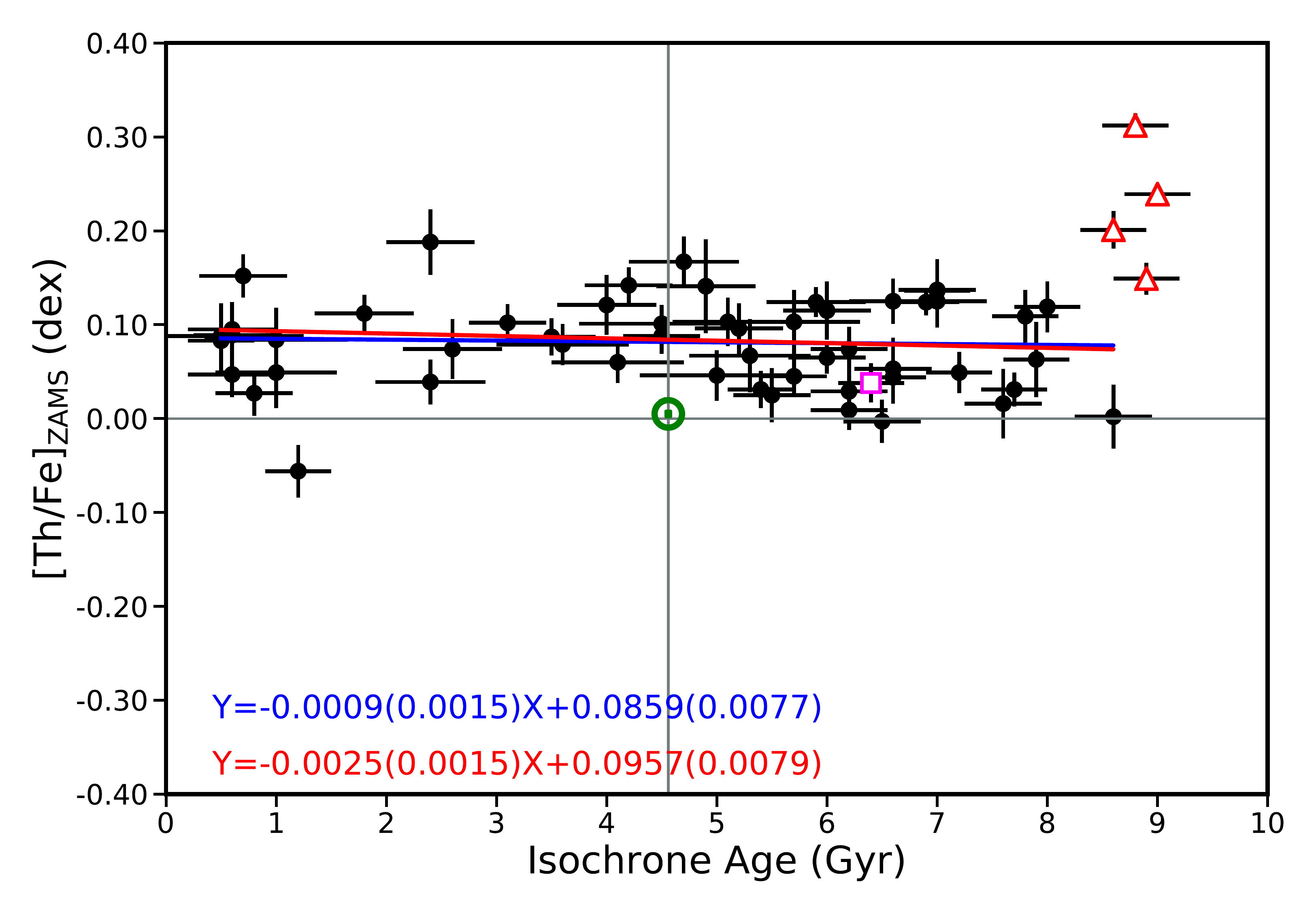}
\includegraphics[width=80mm]{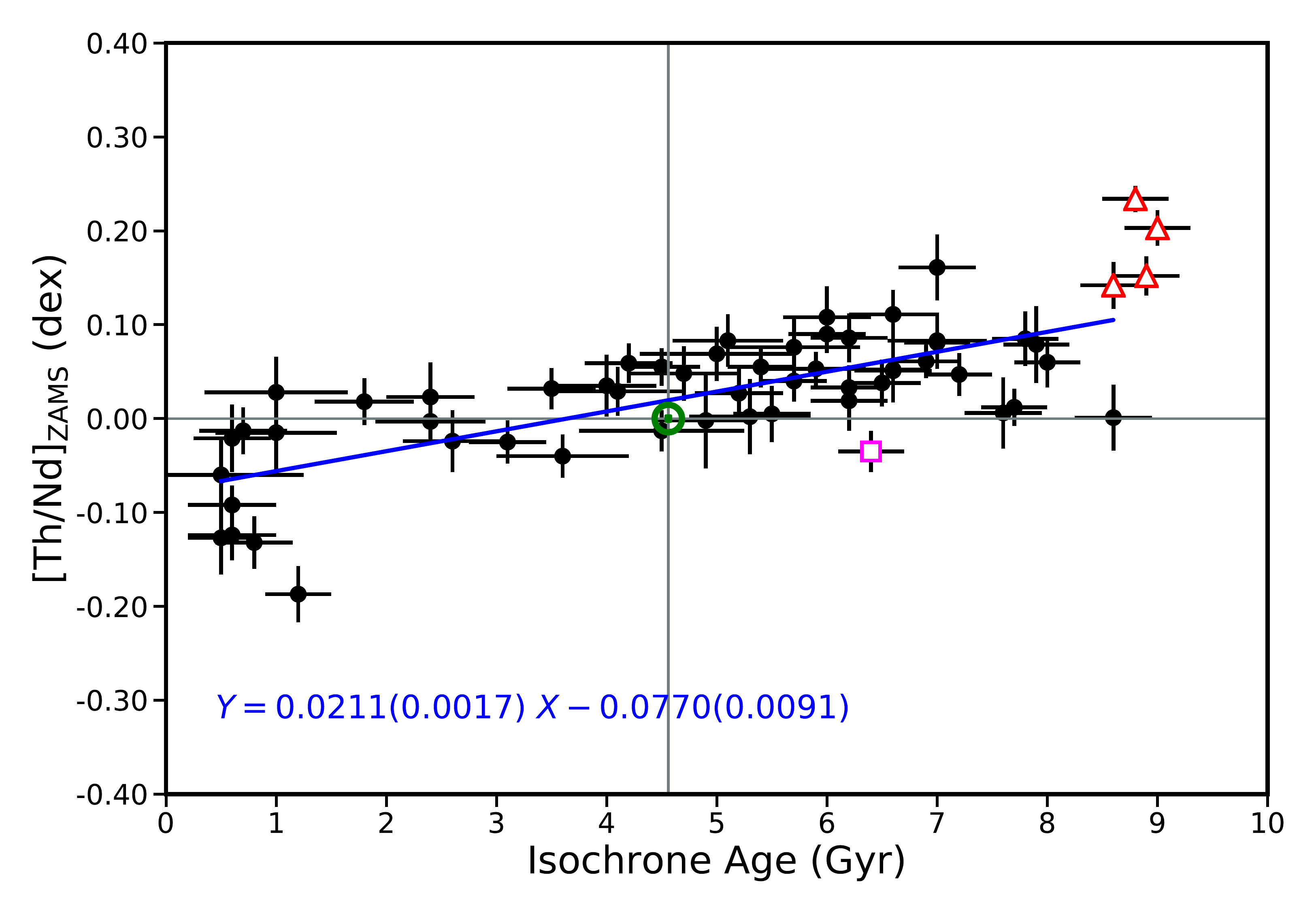}
\includegraphics[width=80mm]{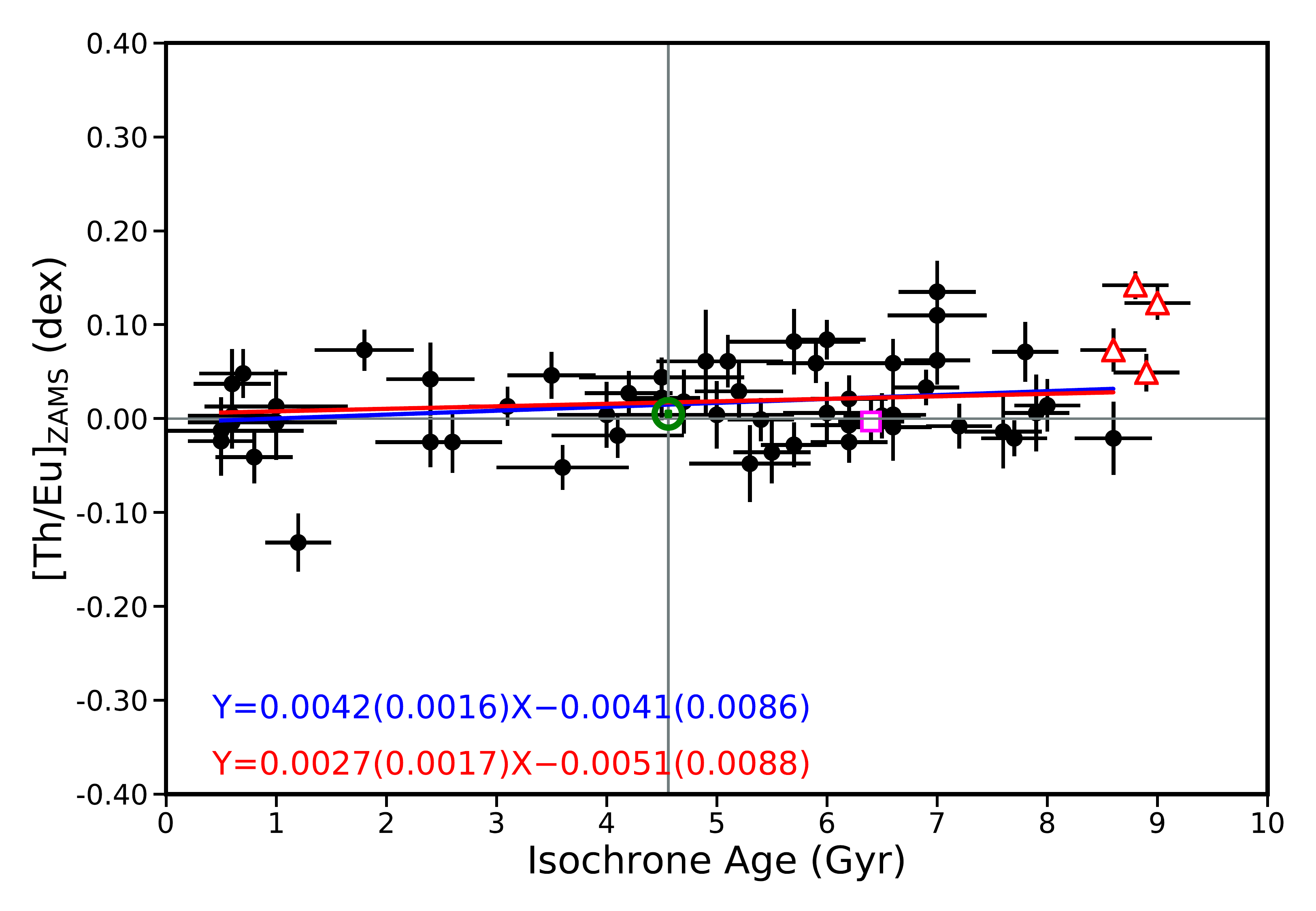}
\includegraphics[width=80mm]{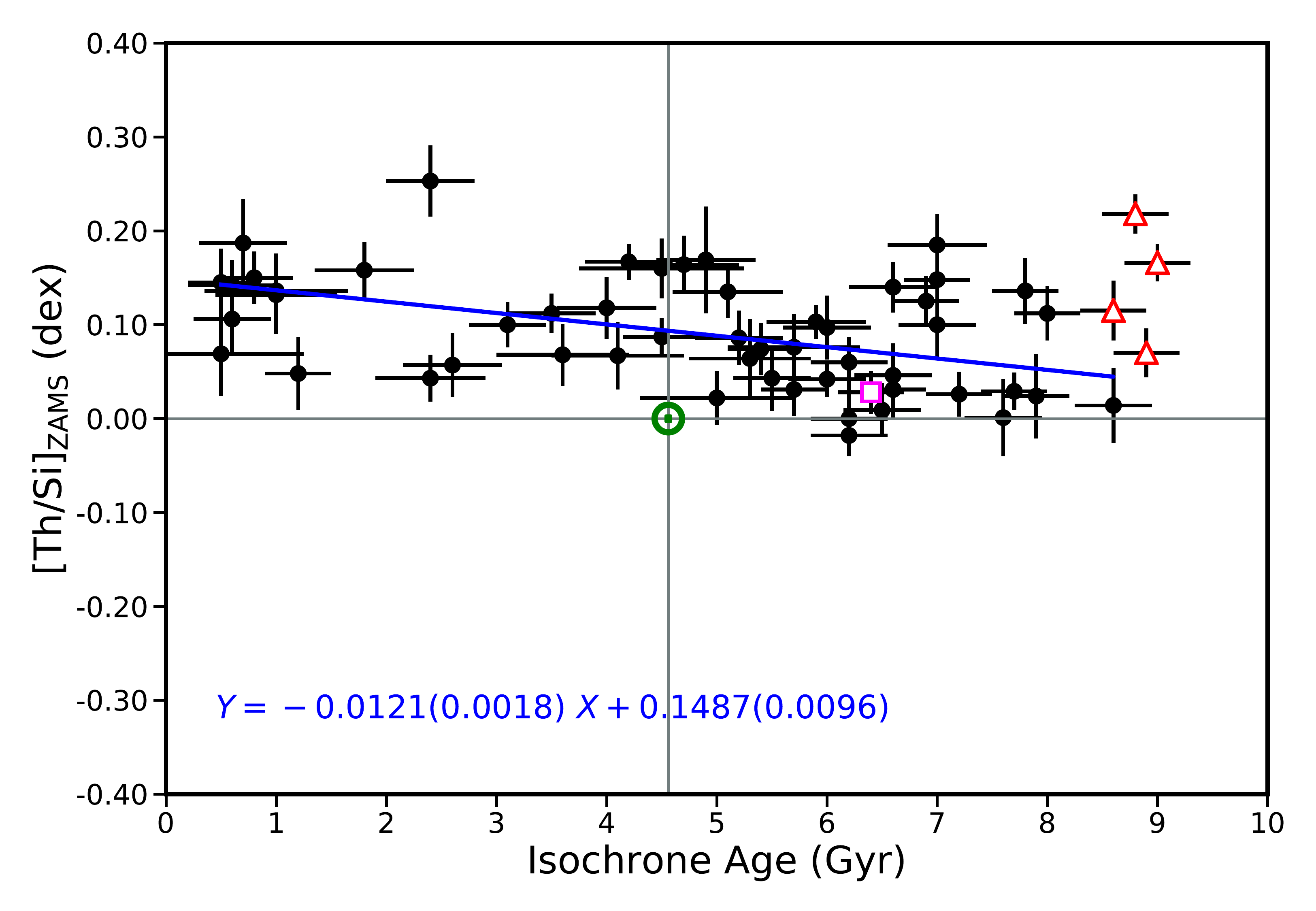}
\caption{
[Th/X]$_{\rm ZAMS}$ as a function of the isochrone stellar age (where X: H, Fe, Nd, Eu and Si):
a statistically robust linear fit is shown in each plot (blue solid line).
The fitting equation with coefficients and their errors inside parenthesis are also displayed.
The fits correspond to 53 thin disc solar twins only (black filled circles).
Specifically for the plots [Th/Fe]$_{\rm ZAMS}$-Age and [Th/Eu]$_{\rm ZAMS}$-Age,
an additional fit is also shown (red solid line) by excluding one borderline 3$\sigma$ outlier
(HIP\,101905, a 1.2\,Gyr old star relatively poor in Th), but the results do not change significantly.
Data of excluded stars are plotted together for illustration purposes only
(red empty triangles: four $\alpha$-rich old stars,
and magenta empty square: the chemically anomalous in $s$-elements HIP\,64150).
Sun's data are also plotted as reference (green solar standard symbol),
although are consistent with the fits.
}
\label{fig_thx_zams_vs_age}
\end{figure*}

The correction we have applied to transform the current values of
[Th/H], [Th/Fe], [Th/Nd], [Th/Eu] and [Th/Si]
into the pristine values on the zero age main sequence (ZAMS)
are those given in \citet{Frebel2007} and \citet{Cayrel2001}
adopting 14.05\,Gyr as half life time of $^{232}$Th ($\tau_{1/2}$(Th)), i.e.
[Th/X]$_{\rm ZAMS}$ = [Th/X]$_{\rm now}$ + (ln(2).log(e)/$\tau_{1/2}$(Th)).($t_{\rm star}$ - $t_{\rm Sun}$),
where [Th/X]$_{\rm ZAMS}$ is the corrected abundance ratio relative to the Sun,
[Th/X]$_{\rm now}$ means the observed current [Th/X], X is a stable element,
$t$ is the star isochrone age in Gyr, and $t_{\rm Sun}$\,=\,4.56\,Gyr.
We have estimated the uncertainties in [Th/X]$_{\rm ZAMS}$
through the propagation of the errors in [Th/X]$_{\rm now}$ and isochrone age as well.

The pristine (corrected) [Th/H]$_{\rm ZAMS}$ correlates well with [Fe/H]
along the short metallicity scale of solar twins,
making [Th/Fe]$_{\rm ZAMS}$ not correlated with [Fe/H]
(see Fig.\ref{fig_thh_thfe_zams_vs_feh},
in which we plot both [Th/H]$_{\rm ZAMS}$ and [Th/Fe]$_{\rm ZAMS}$ as a function of [Fe/H]).

We have fitted [Th/X] as a function of [Fe/H] (observed and ZAMS values for H and Fe) 
and also [Th/X] as a function of isochrone stellar age
(observed and ZAMS values for H and Fe, and ZAMS values only for Nd, Eu and Si)
through a linear fitting approach that minimizes the orthogonal distance of the data points to the fitting curve
thats accounts for the uncertainties in both variables
(it is an orthogonal distance linear regression with variance weighting in both $x$ and $y$).
All fits to the data have been performed
using the Kapteyn \texttt{kmpfit} package \footnote{https://www.astro.rug.nl/software/kapteyn/index.html}.

Figure~\ref{fig_thx_zams_vs_age} shows plots of [Th/X]$_{\rm ZAMS}$ versus isochrone age (X: H, Fe, Nd, Eu and Si).
Table~\ref{tab_fit_results} presents the results of the statistically robust linear fits of [Th/X]$_{\rm observed}$ and [Th/X]$_{\rm ZAMS}$
as a function of both [Fe/H] (where X: H and Fe only) and isochrone age $t$
(i.e. [Th/X]$_{\rm ZAMS}$\,=\,$a$.[Fe/H]\,+\,$b$ and [Th/X]$_{\rm ZAMS}$\,=\,$a$.$t$\,+\,$b$).

The four main results about [Th/H]$_{\rm ZAMS}$ and [Th/Fe]$_{\rm ZAMS}$ in solar twins
as a function of [Fe/H] and isochrone age are as follows.

{\bf (i)} [Th/H]$_{\rm ZAMS}$ correlates very well with [Fe/H]
and the slope of this correlation is smaller than the unity (+0.671$\pm$0.049).
It is not surprise that [Th/Fe]$_{\rm ZAMS}$ decreases with [Fe/H] over the restrict metallicity range of solar twins.
Note that the relative uncertainty of the slope $a$
makes the linear correlations [Th/H]$_{\rm ZAMS}$-[Fe/H] and [Th/Fe]$_{\rm ZAMS}$-[Fe/H] statistically significant
(about 14 and 7\,$\sigma$ of significance, respectively).
See Fig.~\ref{fig_thx_zams_vs_age} and Tab.~\ref{tab_fit_results}.
This suggests that the abundance of Th in the interstellar medium
keeps growing in some way during the Galactic thin disc evolution,
despite the natural radioactive decay of Th and the iron production driven by the contribution from SNIa relatively to SN-II.
We refer the reader to the temporal increase of abundance ratios
relative to iron of stable $r$-process elements such as Nd, Sm, Eu, Gd and Dy,
as reported by \citet{Spina2018} (see Fig.~5 in their paper).

{\bf (ii)} The pristine Th abundance in the sample solar twins
seems to have been super-solar on average during the evolution of Galactic thin disc
and it has certainly become super-solar since the epoch of the Sun formation.
There are no stars with [Th/H]$_{\rm ZAMS}$\,<\,0\,dex with ages smaller than the solar one.
We have obtained for the entire sample an average of [Th/H]$_{\rm ZAMS}$\,=\,+0.080\,dex
with a standard deviation of 0.058\,dex.
It ranges from -0.117 up to +0.257\,dex (from 76 to 181\,per\,cent of the ZAMS solar value),
showing a linear temporal increase from +0.037\,dex 8.6\,Gyr ago up to +0.138\,dex now
(8.6\,Gyr is the isochrone age of the oldest star in our analised sample).
Note that the uncertainty of the slope $a$ of the relation [Th/H]$_{\rm ZAMS}$-Age
corresponds to only 13\,per\,cent of its value (about 8\,$\sigma$ of significance).
See Fig.~\ref{fig_thx_zams_vs_age} and Tab.~\ref{tab_fit_results}.
The Sun at ZAMS seems to be somehow deficient in Th
when compared against twin stars, specially for stars younger than itself.
It is also interesting to note that all four $\alpha$-rich stars
are richer in Th on average than the other Galactic thin disc stars of similar ages;
exactly as found for Nd, Sm and Eu in \citet{Spina2018}.

{\bf (iii)} [Th/Fe]$_{\rm ZAMS}$ shows no correlation with the isochrone age.
Note that the relative error of the slope is greater than 100\,per\,cent.
[Th/Fe]$_{\rm ZAMS}$ has likely kept super-solar during the Galactic thin disc evolution,
showing an average of +0.086$\pm$0.008\,dex ($rms$\,=\,0.047\,dex).
Super-solar Th/Fe ratios for youngest stars could be due to an increased production of Th relatively to Fe,
perhaps due to neutron star mergers.
\citet{Spina2018} also found that the slope of [X/Fe]-Age for stable $n$-capture elements
(in dex.Gyr$^{-1}$ unity) seems to be anti-correlated with the solar $s$-process contribution percentage
taken from \citet{Bisterzo2014} (see Fig.~6 in \citet{Spina2018}),
such that the $r$-process elements Eu, Gd, Dy and Sm
(showing $s$-process contributions smaller than about 35\,per\,cent)
have the smallest slopes in modulus like Th,
which exhibits a statistically null slope
even after removing one borderline 3$\sigma$ threshold outlier
(HIP\,101905 that is a young star relatively poor in Th for its metalllicity).
The slope of [Th/Fe]-Age becomes -0.0025$\pm$0.0015 without HIP\,101905,
which is not significantly different from zero (presenting a confidence level of about 1.7$\sigma$).
See Fig.~\ref{fig_thx_zams_vs_age} and Tab.~\ref{tab_fit_results}.

{\bf (iv)} There are uniform real dispersions of [Th/H]$_{\rm ZAMS}$ and [Th/Fe]$_{\rm ZAMS}$
over the whole isochrone age scale on average ($rms$ close to 0.056 and 0.047\,dex, respectively).

%
%
\begin{table*}
\centering
\caption{
Results of the linear fits of [Th/X] against [Fe/H] and isochrone stellar age $t$:
[Th/X] = $a$(${\pm}{\sigma}_{a}$).[Fe/H] + $b$(${\pm}{\sigma}_{b}$)
(observed and ZAMS [Th/X] values for X: H and Fe only),
and [Th/X]  = $a$(${\pm}{\sigma}_{a}$).$t$ + $b$(${\pm}{\sigma}_{b}$)
(observed and ZAMS [Th/X] values for X: H and Fe, and ZAMS [Th/X] values for X: Nd, Eu and Si).
The slope divided by its error is shown in the last column.
Specifically for [Th/Fe]$_{\rm ZAMS}$-Age and [Th/Eu]$_{\rm ZAMS}$-Age,
results are also presented by excluding one borderline 3$\sigma$ outlier
(HIP\,101905, a young star relatively poor in Th).
These two extra fits are marked with an asterisk (*).
The fitting approach is explained in Sect.~\ref{results}.
}
\label{tab_fit_results}
\begin{tabular}{lrrrrrrr}
\hline
                           & $a$     & $\sigma_{a}$ & $b$    & $\sigma_{b}$ & $rms$ & $\chi^{2}/\nu$ & {\textbar}$a${\textbar}/$\sigma_{a}$ \\
\hline
[Th/X] {\it vs.} [Fe/H]    &         &              &        &              &       &                &  \\
                           &         &              & (dex)  & (dex)        & (dex) &                &  \\
\hline
\null [Th/H]$_{\rm obs}$   & +0.904   & 0.044       & +0.080 & 0.003        & 0.071 &            8.8 & 20.5 \\
\null [Th/Fe]$_{\rm obs}$  & -0.114   & 0.043       & +0.080 & 0.003        & 0.071 &            9.2 &  2.6 \\
\null [Th/H]$_{\rm ZAMS}$  & +0.671   & 0.049       & +0.081 & 0.003        & 0.040 &            2.6 & 13.7 \\
\null [Th/Fe]$_{\rm ZAMS}$ & -0.338   & 0.049       & +0.082 & 0.003        & 0.040 &            2.6 &  6.9 \\
\hline
\null [Th/X] {\it vs.} $t$ &         &              &         &             &       &                &  \\
                           & (dex.Gyr$^{-1}$) & (dex.Gyr$^{-1}$) & (dex)   & (dex)  & (dex) &  &  \\
\hline
\null [Th/H]$_{\rm obs}$    & -0.0353  & 0.0016      & +0.246 & 0.009        & 0.058 &            4.5 & 22.1 \\
\null [Th/Fe]$_{\rm obs}$   & -0.0247  & 0.0015      & +0.195 & 0.008        & 0.048 &            3.5 & 16.5 \\
\null [Th/H]$_{\rm ZAMS}$   & -0.0117  & 0.0015      & +0.138 & 0.008        & 0.056 &            5.2 &  7.8 \\
\null [Th/Fe]$_{\rm ZAMS}$  & -0.0009  & 0.0015      & +0.086 & 0.008        & 0.047 &            3.6 &  0.6 \\
\null *[Th/Fe]$_{\rm ZAMS}$ & -0.0025  & 0.0015      & +0.096 & 0.008        & 0.043 &            3.2 &  1.7 \\
\null [Th/Nd]$_{\rm ZAMS}$  & +0.0211  & 0.0017      & -0.077 & 0.009        & 0.046 &            2.4 & 12.4 \\
\null [Th/Eu]$_{\rm ZAMS}$  & +0.0042  & 0.0016      & -0.004 & 0.009        & 0.043 &            2.4 &  2.6 \\
\null *[Th/Eu]$_{\rm ZAMS}$ & +0.0027  & 0.0017      & -0.005 & 0.009        & 0.039 &            2.1 &  1.6 \\
\null [Th/Si]$_{\rm ZAMS}$  & -0.0121  & 0.0018      & +0.149 & 0.010        & 0.053 &            3.1 &  6.7 \\
\hline
\end{tabular}
\end{table*}

By comparing Th against neodymium (Nd) and europium (Eu), two others $n$-capture elements
that are otherwise stable nuclei indeed,
we have noticed that [Th/Nd]$_{\rm ZAMS}$ decreases with the stellar age,
but [Th/Eu]$_{\rm ZAMS}$ has kept constant around solar ratio over the Galactic thin disc evolution.
Whilst [Th/Nd]$_{\rm ZAMS}$ has a linear decrease from +0.104\,dex 8.6\,Gyr ago up to -0.077\,dex now on average
(fitting $rms$\,=\,0.046\,dex),
the average of [Th/Eu]$_{\rm ZAMS}$ oscillates around +0.014\,dex with a standard deviation of 0.045\,dex.
The slope of the relation [Th/Nd]$_{\rm ZAMS}$-Age has a 12.4$\sigma$ of significance,
and the relation [Th/Eu]$_{\rm ZAMS}$-Age has just 2.4$\sigma$ of significance.
In case of removing one borderline outlier slightly beyond 3$\sigma$
(HIP\,101905 that is a young star relatively poor in Th for its metalllicity),
the slope of [Th/Eu]-Age is even more compatible with zero (1.6$\sigma$ of significance).
See Fig.~\ref{fig_thx_zams_vs_age} and Tab.~\ref{tab_fit_results}.
These results confirm that Th follows Eu during the Galaxy's thin disc evolution. However, it does not follow Nd.
In fact, the contribution of $s$-process for the abundance Nd
in the Solar System is estimated to be 57.5$\pm$4.1\,per\,cent,
while for Eu it is 6.0$\pm$0.4\,per\,cent \citep{Bisterzo2014, Spina2018}.
The decrease of [Th/Nd]$_{\rm ZAMS}$ with time may be explained by an increasing contribution
from the $s$-process due to low-mass AGB stars in the Galactic thin disc as discussed by \citet{Spina2018}.
Therefore, our observations are in line with Nd not being a pure $r$-process element,
but likely having an important $s$-process contribution, as already suggested by \citet{Bisterzo2014}.

On the other hand [Th/Si]$_{\rm ZAMS}$ has likely had a temporal increasing evolution
from +0.045\,dex 8.6\,Gyr ago up to +0.149\,dex now on average (over the super-solar regime).
The anti-correlation [Th/Si]$_{\rm ZAMS}$-Age has a slope with a relative uncertainty of 15\,per\,cent only
or about 7$\sigma$ of significance
(under a $rms$\,=\,0.053\,dex; see Fig.~\ref{fig_thx_zams_vs_age} and Tab.~\ref{tab_fit_results}).
Silicon (Si) is one of the most abundant refractory elements in a terrestrial planet
and its planetary abundance is closely related to the mantle mass \citep{McDonough2003}.
Thus the abundance ratio Th/Si provides comparison of the Th abundance
among telluric planets with different silicate mantle thickness \citep{Unterborn2015}.
A higher [Th/Si]$_{\rm ZAMS}$ can indirectly indicate a higher probability of having 
a convective mantle in a terrestrial planet due to a higher radioactive energy budget from the Th decay.

Thus our results suggest that most solar twins
in the Galactic thin disc have the same capability as the Sun
to have rocky planets with convective mantles, and perhaps suitable for life.

%
\section{Conclusions}
\label{conclusions}

We have confirmed that there is a large energy budget from the Th decay
for keeping the mantle convection and thickness of potential rocky planets
around Sun-like stars, since the Galactic thin disc formation until now,
because we have measured [Th/H]$_{\rm ZAMS}$ as super-solar on average during the Galactic thin disc life.
Whilst the observed [Th/H] varies from -0.117 up to +0.257\,dex
(from 76 up to 181\,per\,cent of the current observed solar value),
[Th/H]$_{\rm ZAMS}$ seems to have linearly changed on average from +0.037\,dex 8.6\,Gyr ago up to +0.138\,dex now
(from 109 up to 137\,per\,cent of the ZAMS solar value),
showing a uniform dispersion of about 0.056\,dex (linear fitting $rms$),
such that [Th/H]$_{\rm ZAMS}$ seems to have certainly become super-solar for all analysed solar twin stars
since the epoch of Solar System formation.
[Th/Fe]$_{\rm ZAMS}$ has kept nearly constant and super-solar as well during the Galactic thin disc evolution
(around +0.086$\pm$0.008\,dex under a linear fitting $rms$ equal to 0.047\,dex).
The Sun at ZAMS actually seems to be deficient in Th
when specifically compared against younger solar twins.
Whilst the observed [Th/Fe]$_{\rm obs}$ is linearly well anti-correlated with isochrone age,
[Th/Fe]$_{\rm ZAMS}$ shows no relation with age.
They both actually seem to have been super-solar on average during practically the entire Galactic thin disc evolution.
[Th/Nd]$_{\rm ZAMS}$ has linearly decreased from +0.104\,dex 8.6\,Gyr ago up to -0.077\,dex now on average.
[Th/Eu]$_{\rm ZAMS}$ has likely kept constant around solar ratio during the Galactic thin disc evolution,
exhibiting a moderate dispersion of about 0.04\,dex.
This implies that Th follows Eu during the Galaxy's thin disc evolution, but it does not Nd.
The decrease of [Th/Nd]$_{\rm ZAMS}$ with time may be explained by
a crescent contribution from the $s$-process in low-mass AGB stars,
indicating that Nd is not exactly a pure $r$-process element.
In fact, it has an estimated $s$-process production close to 60\,per\,cent in the Solar System \citep{Bisterzo2014}.
[Th/Si]$_{\rm ZAMS}$ has probably had a real increase from +0.045\,dex 8.6\,Gyr ago up to +0.149\,dex now on average.

Our results suggest that solar twin stars in the Galactic thin disc are as probable as the Sun
to host rocky planets with convective mantles,
and perhaps with suitable geological conditions for habitability.

\section*{Acknowledgements}

RB acknowledges the CAPES grant (093.875.006-24).
AM acknowledges the CNPq grant (309562/2015-5).
JM thanks FAPESP (2012/24392-2).
The research by MA has been supported by the Australian Research Council
(grants FL110100012 and DP150100250).
We all are grateful for the anonymous referee by his valuable revision to improve the paper.











\bsp	
\label{lastpage}
\end{document}